\documentclass[useAMS,usenatbib]{mn2e}

\usepackage{graphicx}
\usepackage{aas_macros}
\usepackage{amsfonts,amssymb}
\usepackage{color}

\newcommand{\lya}{Ly$\alpha$}

\newcommand{\hi}{H~{\sc i}}

\newcommand{\heii}{He~{\sc ii}}

\newcommand{\taueffa}{$\tau_{\rm eff}^{\alpha}$}

\newcommand{\Tdelta}{$T(\bar{\Delta})$}
\newcommand{\Tbar}{${T(\bar{\Delta})}$}

\newcommand{\G}{$\Gamma$}
\newcommand{\Nion}{$\dot{N}_{\rm ion}$}
\newcommand{\mfp}{$\lambda_{912}$}
\newcommand{\eps}{$\epsilon_{912}$}
\newcommand{\epsgal}{$\epsilon_{912}^{\rm G}$}
\newcommand{\epsfifteen}{$\epsilon_{1500}^{\rm G}$}

\title[UV Background over $2 < z < 5$]{New Measurements of the Ionizing Ultraviolet Background over {\boldmath $2 < z < 5$} and Implications for Hydrogen Reionization}

\author[Becker \& Bolton]
   {George D.~Becker$^1$\thanks{gdb@ast.cam.ac.uk} \& James S.~Bolton$^2$ \\
   $^1$Kavli Institute for Cosmology and Institute of Astronomy, University of Cambridge, Madingley Rd, Cambridge, CB3 0HA, UK \\
   $^2$School of Physics and Astronomy, University of Nottingham, University Park, Nottingham NG7 2RD}
   \date{Draft version \today}
   
\begin{document}

\label{firstpage}

\maketitle

\begin{abstract}  

We present new measurements of the intensity of the ionizing
ultraviolet background and the global emissivity of ionizing photons
over $2 < z < 5$.  Our results are based on a suite of updated
measurements of physical properties of the high-redshift intergalactic
medium (IGM), including gas temperatures and the opacity of the IGM to
\lya\ and ionizing photons.  Consistent with previous works, we find a
relatively flat hydrogen photoionization rate over $2 < z < 5$,
although our measurements are roughly a factor of two higher than the
2008 values of Faucher-Gigu{\`e}re et~al., due primarily to our lower
gas temperatures.  The ionizing emissivity we derive is also generally
higher than other recent estimates due to a combination of lower gas
temperatures, higher ionizing opacity, and an accounting of
cosmological radiative transfer effects.  We find evidence that the
emissivity increases from $z \sim 3$ to 5, reaching $\sim$5 ionizing
photons per atom per gigayear at $z = 4.75$ for realistic galaxy
spectra.  We further find that galaxies must dominate the emissivity
near 1 Ryd at $z \ge 4$, and possibly at all redshifts $z \ge 2.4$.
Our results suggest that the globally-averaged ionizing ``efficiency''
of star-forming galaxies increases substantially with redshift over
$3.2 \le z \le 4.75$.  This trend is consistent with the conclusion
often drawn from reionization models that the ionizing efficiency of
galaxies must be higher during reionization in order for galaxies to
reionize the IGM by $z=6$.  Our emissivity values at $z \sim 5$
suggest that ionizing photons may have been a factor of two more
abundant during the final stages of reionization than previously
indicated.  The evolution of the ionizing emissivity over $2 < z < 5$
suggests, moreover, that the steep decline in the photoionization rate
from $z \sim 5$ to 6 may indicate a rapid evolution in the mean free
path at $z > 5$.

\end{abstract}


\section{Introduction}

The ionizing ultraviolet background (UVB) is a key probe of the
interaction between luminous sources and the intergalactic medium
(IGM).  Following hydrogen reionization, the nearly complete
ionization of the IGM is maintained by ultraviolet photons from
star-forming galaxies and active galactic nuclei (AGN).  Many (or
most) of these sources are too faint, however, to observe directly at
high redshifts.  Moreover, the ionizing portions of their spectra are
largely obscured due to absorption from the IGM.  Measurements of the
UVB derived from the properties of the IGM itself therefore provide
unique insights into critical aspects of galaxy and AGN evolution that
are difficult to study by other means.

Studies of the UVB have particular relevance for understanding how and
when the IGM became reionized.  For example, the properties of the UVB
following reionization provide a boundary condition on the evolution
of the ionizing radiation field during reionization itself.  The
intensity of the UVB at $z \sim 5-6$, in particular, can help to
determine how abundant ionizing photons in the IGM may have been
during the final stages of reionization.  Previous measurements of a
low ionizing emissivity at these redshifts suggested that reionization
may have been a ``photon-starved'' process, with only $\sim 2-3$
ionizing photons being emitted per baryon per gigayear at $z = 6$
\citep{me2003,meiksin2005,bolton2007,kuhlen2012}.

The redshift evolution of the UVB can also help to identify trends
that clarify how galaxies and/or AGN were able to produce and emit
enough ionizing photons to reionize the IGM by $z = 6$
\citep{fan2006b}.  Direct searches for star-forming galaxies at $z >
6$ indicate that the global star formation rate density declines
steeply with redshift over $6 < z < 12$
\citep[e.g.,][]{ouchi2009,castellano2010,bunker2010,bouwens2006,bouwens2008,bouwens2011a,bouwens2012,oesch2010a,mclure2010,mclure2013,schenker2013,ellis2013}.
Even when accounting for sources below the detection limits,
reionization models based on the measured non-ionizing UV galaxy
luminosity functions typically require that reionization-era galaxies
emit significantly more ionizing photons, relative to their
non-ionizing luminosity, than their lower-redshift counterparts
\citep[e.g.,][]{ouchi2009,kuhlen2012,mitra2013,robertson2013,ferrara2013}.
This ionizing ``efficiency'' is often parametrized in terms of the
relative escape fraction of ionizing photons, but also depends on a
galaxy's intrinsic UV spectral energy distribution.

Direct searches for ionizing emission from star-forming galaxies
suggests that typical ionizing escape fractions (and/or intrinsic
ratios of ionizing to non-ionizing luminosity) may indeed be larger at
earlier times.  \citet{siana2010}, for example, find $f_{\rm esc} <
0.04$ at $z \sim 1$, while \citet{nestor2013} find $f_{\rm esc} \sim
5-7$ per cent for Lyman break galaxies (LBGs) and $f_{\rm esc} \sim
10-30$ per cent for fainter Lyman-alpha emitters (LAEs) \citep[see
  also][]{mostardi2013}.  Although directly detecting the ionizing
flux from galaxies becomes extremely difficult at $z \ge 4$, indirect
evidence that the escape fraction may continue increasing with
redshift at $z > 3$ comes from an observed increase with redshift in
the fraction of LBGs showing \lya\ emission
\citep{stark2010,stark2011} combined with an anti-correlation between
\lya\ fraction and the covering fraction of \hi\ gas
\citep{jones2012}.

Measurements of the UVB can help clarify these trends by providing a
census of the ionizing emissivity from all sources, a quantity that is
essentially impossible to obtain from direct observations alone.  The
ionizing emissivity can then be compared to the non-ionizing UV
emissivity measured directly from galaxies and AGN in order to assess
the evolution in ionizing efficiency.

In this paper we present new calculations of the \hi\ photoionization
rate and the ionizing emissivity over $2 < z < 5$. Our results are
based on multiple recent measurements of the physical properties of
the high-redshift IGM, including gas temperatures and the mean opacity
of the IGM to \lya\ and ionizing photons.  Our work builds upon
previous measurements of the UVB using the mean \lya\ opacity
\citep[e.g.,][]{rauch1997,songaila1999,mcdonald2001a,meiksin2003,tytler2004,jena2005,kirkman2005,bolton2005,fan2006b,bolton2007,fg2008c,wyithe2011,kuhlen2012}
and the quasar proximity effect
\citep[e.g.,][]{bdo1988,scott2000,dallaglio2008,calverley2011}.  The
new measurements of the IGM properties are significantly more precise
than previous estimates, and allow us to conduct a detailed
examination of the evolution of the UVB in the post-reionization
epoch.  A key advantage of the present work is the fact that we have
self-consistent measurements of the temperature, \lya\ opacity, and
ionizing opacity extending over $2 < z < 5$, allowing us to study the
evolution of the UVB over a wide redshift range and as close as
possible to the reionization epoch itself.

In Section~\ref{sec:data} we briefly describe the IGM measurements
upon which our analysis is based.  The hydrodynamical simulations used
to calibrate these measurements are described in
Section~\ref{sec:simulations}.  In Section~\ref{sec:Gamma} we present
our results for the \hi\ photoionization rate, which are derived from
the \lya\ opacity and gas temperature of the IGM.  In
Section~\ref{sec:emissivity} we combine these results with
measurements of the opacity of the IGM to ionizing photons to derive
the ionizing emissivity.  We then turn towards disentangling the
contribution to the emissivity from AGN and galaxies in
Section~\ref{sec:sources}, and compare our results for the integrated
ionizing emissivity from galaxies to the non-ionizing UV emissivity
measured from galaxy surveys.  We then discuss the implications of our
results for the evolution of the ionizing efficiency of galaxies and
for hydrogen reionization in Section~\ref{sec:discussion} before
concluding in Section~\ref{sec:summary}.  Our results assume a flat
$\Lambda$CDM cosmology with $\Omega_{\rm m} = 0.308$, $\Omega_\Lambda
= 0.692$, $\Omega_{\rm b}h^2 = 0.0222$, $h=0.678$, $\sigma_8 = 0.829$,
and $n_{\rm s} = 0.961$, consistent with recent {\it
  Planck}+WP+highL+BAO constraints from \citet{planckXVI}, although
uncertainties in the cosmological parameters are also taken into
account.

\section{The Data}\label{sec:data}

Our analysis of the UV background draws on multiple recent
measurements of physical properties of the high-redshift IGM.  Here we
briefly describe these measurements, while in Sections~\ref{sec:Gamma}
and \ref{sec:emissivity} we detail how they are used to calculate the
hydrogen photoionization rate and the ionizing emissivity.

The calculations of the hydrogen photoionization rate, \G, are based
on the mean opacity of the IGM to \lya\ photons and the temperature of
the IGM.  The \lya\ opacity is quantified in terms of an effective
optical depth, $\tau_{\rm eff}^{\alpha} = -\ln{\langle F \rangle}$,
where $F$ is the continuum-normalized transmitted flux in the
\lya\ forest.  Measurements of \taueffa\ over $2.15 \le z \le 4.85$
are taken from \citet{becker2013}.  This work used composites of
quasar spectra drawn from the Sloan Digital Sky Survey
\citep{york2000} to perform a differential measurements of the mean
transmitted \lya\ flux as a function of redshift, normalizing the
results to measurements at $z \le 2.5$ made from high-resolution data
by \citet{fg2008a} .  This approach has the advantage of avoiding the
need to continuum-fit individual spectra at high redshifts, where the
\lya\ forest becomes heavily absorbed, enabling a precise measurement
of \taueffa\ up to $z \sim 5$.

Our temperature measurements are taken from \citet{becker2011a}, who
used the curvature of the \lya\ forest in a large set of
high-resolution quasar spectra to determine IGM temperatures over $2.0
\le z \le 4.8$.  At a given redshift, they measured only the
temperature at an optimal overdensity probed by the forest,
$T(\bar{\Delta})$, where the temperature at that density, as
determined from the curvature, did not depend on the overall shape of
the temperature-density relation.  This approach provides an accurate
measurement of the temperature at a single density, but does not
constrain the overall shape of the temperature-density relation.  We
therefore quantify systematic uncertainties in our estimate of the
ionization rate related to the shape of the temperature-density
relation in Section~\ref{sec:Gamma}.  The error bars reported by
\citet{becker2011a} are statistical only; however, they also note a
systematic uncertainty related to the integrated thermal history.  To
account for this, we add 2000~K to the errors in $T(\bar{\Delta})$ at
each redshift, which is close to the typical systematic error.

Translating our values for the ionization rate into estimates of the
ionizing emissivity requires quantifying the opacity of the IGM (and
circum-galactic medium; CGM) to ionizing photons.  For this we combine
two sets of recent measurements.  The first is a set of direct
determinations of the mean free path for 1 Ryd photons, \mfp, over
$2.4 \le z \le 4.9$ derived from composite quasar spectra
\citep[][Worseck et al., in
  prep]{prochaska2009b,omeara2013,fumagalli2013}.  This method has the
advantage that it directly probes the combined IGM+ICM opacity without
needing to know the detailed \hi\ column density distribution.  As
described in the appendix, however, a correct calculation of the
emissivity at $z \simeq 2-3$, where the mean free path is substantial,
requires quantifying the opacity to ionizing photons as a function of
frequency and redshift.  We do this by combining the direct
measurements of \mfp\ with measurements of the incidence rate of
optically-thick Lyman limit systems (LLSs) over $2.7 \le z \le 4.4$
from \citet{songaila2010}.  As described in
Section~\ref{sec:emissivity}, the $n_{\rm LLS}$ measurements are not
used to refine the measurement of \mfp, but rather to approximately
constrain the shape of the \hi\ column density distribution over the
column density range that dominates the ionizing opacity.  This then
enables us to calculate the IGM+CGM opacity as a function of frequency
and redshift.

\section{Hydrodynamical Simulations}\label{sec:simulations}

We use hydrodynamical simulations of the IGM to calibrate our
measurements of the \hi\ ionization rate.  The simulations used here
build on the runs described in \citet{becker2011a} and are performed
with the cosmological hydrodynamical code GADGET-3, an updated version
of the publicly available code GADGET-2 \citep{springel2005}.
Briefly, our primary simulations are performed in a 10$h^{-1}$
comoving Mpc box with a gas particle mass of $9.2\times 10^{4}h^{-1}
M_{\odot}$.  The simulations explore a wide range of IGM thermal
histories which are summarized in Table 2 of \citet{becker2011a}.  In
this work, we supplement these simulations with several additional
runs which we now turn to describe briefly here.

The fiducial cosmology assumed in our simulations consists of a flat
$\Lambda$CDM cosmology with $\Omega_{\rm m} = 0.26$, $\Omega_\Lambda =
0.74$, $\Omega_{\rm b}h^2 = 0.023$, $h=0.72$, $\sigma_8 = 0.80$, and
$n_{\rm s} = 0.96$.  As discussed in \citet{bolton2005} and
\citet{bolton2007}, however, the calibration of \hi\ ionization rate
as a function of \lya\ opacity will depend on cosmology.  We therefore
performed an addition run, C15$_{\rm P}$, with the same thermal
history as run C15 in \citet{becker2011a} but using $\Omega_{\rm m} =
0.308$, $\Omega_\Lambda = 0.692$, $\Omega_{\rm b}h^2 = 0.0222$,
$h=0.678$, $\sigma_8 = 0.829$, and $n_{\rm s} = 0.961$, consistent
with the {\it Planck}+WP+highL+BAO constraints from \citet{planckXVI}.
Corrections to \G\ for cosmology as a function of redshift were then
computed by comparing the results obtained using run C15 and C15$_{\rm
  P}$.  The result was a 14 to 19 per cent reduction in \G\ when using
the {\it Planck} cosmology, a similar but somewhat smaller correction
than that computed from the scaling relations for individual
cosmological parameters derived in \citet{bolton2005} and
\citet{bolton2007}.

Convergence of \G\ with box size and mass resolution were checked
using the ``R'' runs from \citet{becker2011a}, in which we alternately
varied the mass resolution and box size.  The results indicated that
we are well converged with mass resolution at our nominal value of
$M_{\rm gas} = 9.2 \times 10^4~h^{-1}\,{\rm M_{\odot}}$, with at most
a 2 per cent increase in \G\ when using a factor of eight coarser
resolution (run R1).  We did find, however, that the results using our
nominal box size of $10~h^{-1}$Mpc over-predicted \G\ by roughly 10
percent compared to those using the $40~h^{-1}$Mpc box (run R4).  This
is presumably due to the fact that the smaller box contains fewer
rare, deep voids, which tend to decrease the mean \lya\ opacity.  We
therefore applied a redshift-dependent box size correction, finding
that $\Gamma^{40}/\Gamma^{10} = 0.90 -0.018z + 0.0041z^{2}$ provided a
good fit over $2 < z < 5$.  Our convergence results are consistent
with those presented by \citet{boltonbecker2009}.

Our fiducial results were calculated using the density and peculiar
velocity fields from a simulation in which hydrogen reionization
begins at $z = 12$ and is initially heated to a maximum temperature of
$\sim$9,000 K by $z \simeq 9$ (see Figure~\ref{fig:Thist}).  Different
thermal histories will alter the small-scale density field and
peculiar velocity fields via Jeans smoothing, which can potentially
impact the mean \lya\ opacity.  As described in
Appendix~\ref{app:jeans}, however, these effects were tested and found
to have a minimal impact on our results.

\section{\hi\ Ionization Rate}\label{sec:Gamma}

\subsection{Method}\label{sec:Gamma_method}

The intensity of the ionizing ultraviolet background is typically
characterized by either the specific intensity at the Lyman limit,
$J_{912}$, or the total hydrogen ionization rate, $\Gamma$.  These
quantities are related as
\begin{equation}\label{eq:Gamma}
\Gamma(z) = 4\pi \int_{\nu_{912}}^{\infty} \frac{d\nu}{h\nu}J_{\nu}(z)\sigma_{\rm H\, I}(\nu) \, ,
\end{equation} 
where $\sigma_{\rm H\, I}(\nu)$ is the photoionization cross section.
Since the shape of the ionizing background is not known a priori, the
photoionization rate is a more model-independent measure of the
intensity of the UVB.  Within the context of a given cosmological
model, $\Gamma$ can be derived from the opacity of the IGM to
\lya\ photons, provided the gas temperature is also known.  In
photoionization equilibrium, the local optical depth of the IGM to
\lya\ scattering \citep{gunn1965}, neglecting the effects of peculiar
velocities and thermal broadening, is given by
\citep[e.g.,][]{mcdonald2001a}
\begin{equation}\label{eq:tau_lya}
\tau \propto \frac{(1+z)^6 (\Omega_{\rm b} h^2)^2}{T^{0.7}H(z)\Gamma(z)}\Delta^{\beta} \, ,
\end{equation}
where $\Delta = \rho/\langle \rho \rangle$ is the fractional
overdensity.  The temperature scaling here reflects the temperature
dependence of the recombination rate.  For a power law
temperature-density relation of the form $T(\Delta) =
\bar{T}(\Delta/\bar{\Delta})^{\gamma-1}$, the slope with density is
given by $\beta = 2-0.7(\gamma-1)$.

To predict \taueffa\ as a function of \G, the local optical depths
must be integrated over a realistic density distribution, and both
peculiar velocities and thermal broadening must be taken into account.
To do this, we follow \citet{bolton2005} and \citet{bolton2007} in
calibrating our $\Gamma$ results using artificial \lya\ forest spectra
drawn from hydrodynamical simulations.  The simulated spectra are used
to predict \taueffa\ for a given ionization rate and
temperature-density relation.  We thus measure \G\ at a given redshift
by matching the temperatures in the simulation to our observed values
and tuning the ionization rate such that the predicted
\taueffa\ matches the observed value \citep[e.g.,][]{theuns1998}.

The effective optical depths from \citet{becker2013} consist of
twenty-eight \taueffa\ values over $2.15 \le z_{\alpha} \le 4.85$ in
bins of $\Delta z_{\alpha} = 0.1$, while the temperature measurements
consist of eight \Tdelta\ measurements over $2.0 < z_{\rm T} < 4.8$ in
bins of $\Delta z_{\rm T} = 0.4$.  Our approach for determining the
best-fitting values of $\Gamma(z)$ attempts to make optimal use of
these data while accounting for the fact they have different redshift
binnings.

We first determine the best-fitting values of \G\ for each $\tau_{\rm
  eff}^{\alpha}$ measurement using the \Tdelta\ value linearly
interpolated onto redshift $z_{\alpha}$ and a fixed value for
$\gamma$.  For each $z_{\alpha}$ we identify the four nearest
simulation output redshifts, two at $z_{\rm sim} < z_{\alpha}$ and two
at $z_{\rm sim} > z_{\alpha}$.  Using the optical depths along 1000
random lines of sight generated from the native temperature, density,
and velocity fields of the fiducial simulation, we determine the
ionization rate needed to reproduce the observed effective optical
depth at each of the four simulation redshifts.  We adjust these
\G\ values to the interpolated value of \Tdelta\ and chosen value of
$\gamma$ using the scaling relations described in
Appendix~\ref{app:scaling}.  The best-fitting value for \G\ at
$z_{\alpha}$ is then determined using a power law fit to \G\ versus
$1+z_{\rm sim}$.  Cosmology and box size corrections are applied.
Finally, in order to minimize the correlation introduced by
interpolating the temperatures, we average the \G\ results over the
temperature redshift bins.  There are four \taueffa\ measurements per
temperature bin for all except the highest-redshift point, for which
there are three.

Statistical errors for \G\ are calculated using a Monte Carlo approach
in which the values of \taueffa\ and \Tdelta\ are varied randomly.
Realizations for \taueffa\ were drawn using the full covariance matrix
given by \citet{becker2013}.  We also propagate errors in the
cosmological parameters through to our \G\ results by generating
random sets of cosmological parameters from the full {\it Planck}
posterior distributions.  The corresponding errors in \G\ are then
estimated using the scaling relations in \citet{bolton2005} and
\citet{bolton2007}.

The temperature measurements of \citet{becker2011a} specify the
temperature at a specific overdensity at which $T(\bar{\Delta}$) could
be determined independently from the shape of the temperature-density
relation, and do not provide constraints on $\gamma$.  We therefore
evaluate \G\ over a range in $\gamma$ that attempts to bracket the
likely values at these redshifts.  In photoionization equilibrium,
$\gamma$ is expected to asymptotically reach a maximum value of
$\gamma \simeq 1.6$ \citep{huignedin1997}, which we take as an upper
limit.  For our minimum value we adopt $\gamma = 1.2$, which
corresponds to a significantly flattened temperature-density relation.
For simplicity, we adopt $\gamma = 1.4$ as our fiducial value.

In principle $\gamma$ can be lower than 1.2, with the
temperature-density relation becoming flat or even inverted ($\gamma <
1$) immediately following a reionization event.  After reionization,
however, the voids are expected to cool rapidly via adiabatic
expansion, driving $\gamma$ towards its asymptotic value.  Simulations
of hydrogen reionization suggest that $\gamma$ should only be less
than 1.2 near $z \sim 5$, and then only if a majority of the voids are
reionized close to $z=6$ \citep[e.g.,][]{trac2008,furlanetto2009}, a
scenario currently disfavored by observations of $z=6.6$ \lya-emitting
galaxies \citep[e.g.,][]{ouchi2010}.  In a patchy and extended helium
reionization \citep[e.g.,][]{becker2011a}, the globally-averaged
temperature-density relation is also unlikely to flatten dramatically
\citep{mcquinn2009a,compostella2013}, although it may be flat or
inverted locally.  We note that some analyses of the \lya\ flux
probability distribution function have yielded tentative evidence of a
flat or inverted temperature-density relation near $z\sim 2-3$
\citep[e.g.,][]{bolton2008,viel2009}, although it has been argued
these measurements are sensitive to systematic effects such as quasar
continuum placement \citep{lee2012}, metal contamination and
underestimated jack-knife error bars \citep{rollinde2013}.  A recent
measurement based on the cutoff in Doppler parameters of \lya\ forest
lines by \citet{rudie2012} favors $\gamma \simeq 1.5$ at $z = 2.4$.
We also note that a simple power-law parametrization for the
temperature-density relation is unlikely to be sufficient during or
immediately following reionization, as significant scatter and a
departure from a single power law are expected
\citep[e.g.,][]{bolton2004,trac2008,furlanetto2009,mcquinn2009a,meiksin2012,compostella2013}.
Nevertheless, the scaling relations in Appendix~\ref{app:scaling} can
be used to extend our results to flatter temperature-density
relations.

\begin{table*}
   \renewcommand{\arraystretch}{1.3}
   \caption{Results and error budget for $\log{\Gamma}$.  Units for
     \G\ are $10^{-12}\,{\rm s^{-1}}$.  The nominal values are for
     $\gamma = 1.4$.  Statistical errors reflect the diagonal terms of
     the covariance matrix only.}
   \vspace{-6pt}
   \label{tab:Gamma_errors}
   \begin{center}
   \begin{tabular*}{\textwidth} {@{\extracolsep{\fill}}l*{7}{c}}
   \hline
   z                              &         2.40  &        2.80  &        3.20  &        3.60  &        4.00  &        4.40  &        4.75  \\
   \hline
   Nominal value                  &        0.015  &    $-$0.066  &    $-$0.103  &    $-$0.097  &    $-$0.072  &    $-$0.019  &    $-$0.029  \\
   \hline
   \taueffa\ errors only          &   $\pm$0.040  &  $\pm$0.026  &  $\pm$0.018  &  $\pm$0.013  &  $\pm$0.012  &  $\pm$0.015  &  $\pm$0.026  \\
   $T(\bar{\Delta})$ errors only  &   $\pm$0.016  &  $\pm$0.017  &  $\pm$0.021  &  $\pm$0.028  &  $\pm$0.035  &  $\pm$0.043  &  $\pm$0.056  \\
   Cosmology errors only          &   $\pm$0.019  &  $\pm$0.019  &  $\pm$0.020  &  $\pm$0.021  &  $\pm$0.022  &  $\pm$0.023  &  $\pm$0.024  \\
   Total statistical error        &   $\pm$0.047  &  $\pm$0.037  &  $\pm$0.036  &  $\pm$0.038  &  $\pm$0.043  &  $\pm$0.052  &  $\pm$0.070  \\
   $\gamma = 1.6$                 &     $+$0.084  &    $+$0.086  &    $+$0.083  &    $+$0.077  &    $+$0.071  &    $+$0.066  &    $+$0.064  \\
   $\gamma = 1.2$                 &     $-$0.084  &    $-$0.086  &    $-$0.083  &    $-$0.077  &    $-$0.071  &    $-$0.066  &    $-$0.064  \\
   Jeans smoothing                &   $_{-0.016}^{+0.001}$  &  $_{-0.008}^{+0.006}$  &  $_{-0.003}^{+0.012}$  &  $_{-0.003}^{+0.016}$  &  $_{-0.004}^{+0.021}$  &  $_{-0.005}^{+0.023}$  &  $_{-0.014}^{+0.023}$  \\
   Total systematic error         &   $_{-0.100}^{+0.085}$  &  $_{-0.095}^{+0.092}$  &  $_{-0.085}^{+0.095}$  &  $_{-0.080}^{+0.094}$  &  $_{-0.075}^{+0.092}$  &  $_{-0.071}^{+0.089}$  &  $_{-0.077}^{+0.086}$  \\
   Total error                    &   $_{-0.146}^{+0.132}$  &  $_{-0.131}^{+0.129}$  &  $_{-0.121}^{+0.130}$  &  $_{-0.118}^{+0.131}$  &  $_{-0.117}^{+0.135}$  &  $_{-0.122}^{+0.140}$  &  $_{-0.147}^{+0.156}$  \\
   \hline
   \end{tabular*}
   \end{center}
\end{table*}

\subsection{Photoionization rate results}\label{sec:Gamma_results}

\begin{figure}
   \begin{center}
   \includegraphics[width=0.45\textwidth]{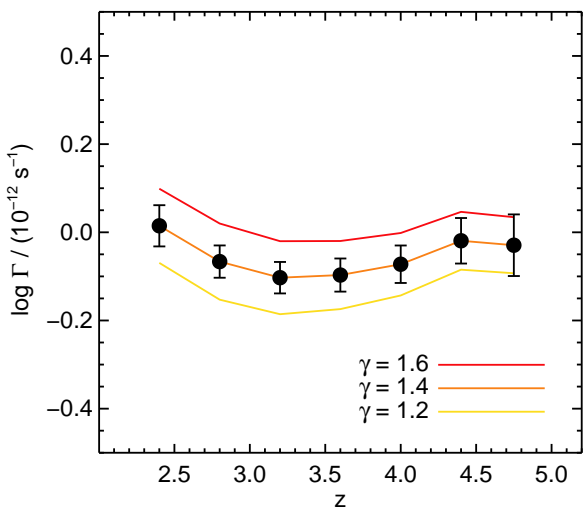}
   \vspace{-0.05in}
   \caption{The hydrogen ionization rate as a function of redshift.
     Solid points with 1$\sigma$ statistical error bars give the
     best-fitting \G\ for our fiducial parameter for the
     temperature-density relation ($\gamma$ = 1.4).  Solid lines give
     \G\ for $\gamma = [1.6,1.4,1.2]$ (top to bottom).  Additional,
     smaller systematic uncertainties related to the density field are
     summarized in Table~\ref{tab:Gamma_errors}.}
   \label{fig:G12}
   \end{center}
\end{figure}

Our results for the \hi\ photoionization rate are shown in
Figure~\ref{fig:G12}, with the error budget is summarized in
Tables~\ref{tab:Gamma_errors} and \ref{tab:Gamma_covar}.  We find
$\Gamma(z)$ to be nearly flat (to within $\sim$0.1~dex) over $ 2 < z <
5$.  There is mild evidence for evolution with redshift, but it is not
highly significant given the correlated statistical errors in
$\Gamma(z)$ (Table~\ref{tab:Gamma_covar}) Systematic errors arising
from the uncertainty in $\gamma$ dominate the error budget for all but
the highest-redshift bin.  Errors arising from the unknown degree of
Jeans smoothing, in contrast, are relatively small at all redshifts
(see Table~\ref{tab:Gamma_errors}).

\begin{figure*}
   \centering
   \begin{minipage}{\textwidth}
   \begin{center}
   \includegraphics[width=1.00\textwidth]{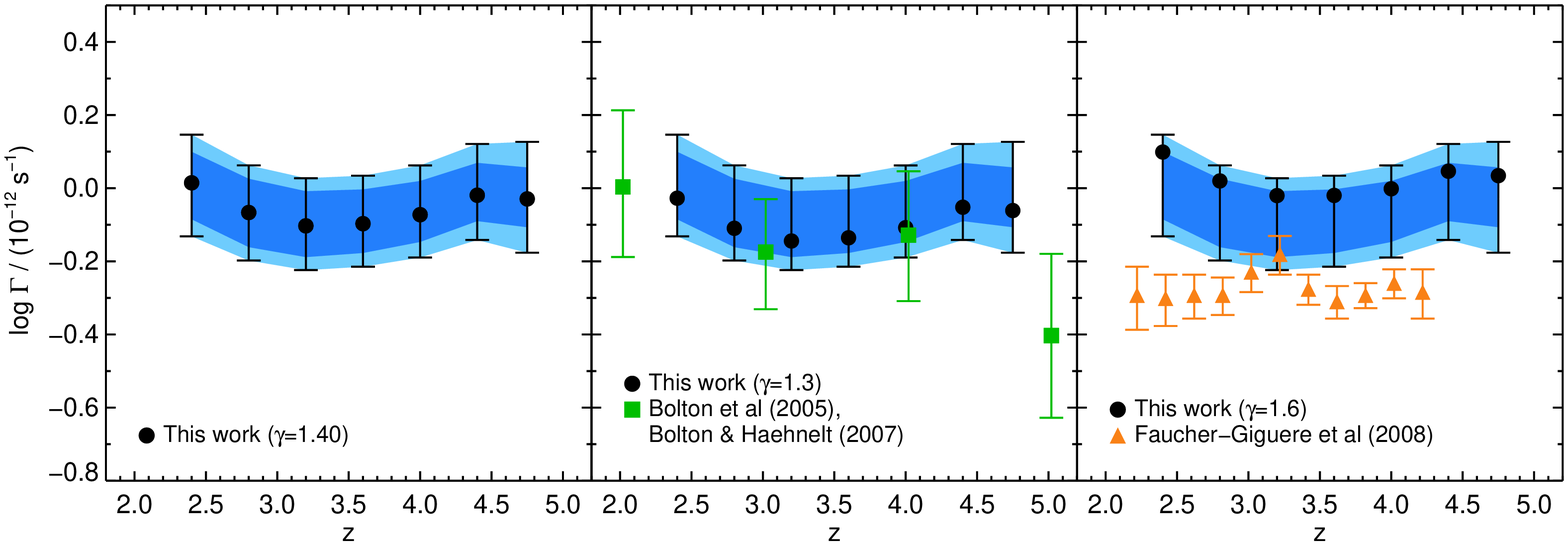}
   \vspace{-0.05in}
    \caption{The hydrogen ionization rate, \G, along with literature
      values over $2 < z < 5$.  In each panel the inner shaded band
      gives the total range of systematic uncertainty, while the outer
      shaded band gives the total statistical error.  In the left-hand
      panel the filled circles show the nominal values of \G\ for our
      fiducial model with $\gamma = 1.4$.  In the middle and left-hand
      panels, the filled circles show \G\ for values of $\gamma$
      corresponding to those adopted in previous works
      \citep{bolton2005,bolton2007,fg2008c}.  Literature values have
      been adjusted for cosmology.  Note that the error bars for
      \citet{bolton2005} and \citet{bolton2007} include uncertainties
      in $\tau_{\rm eff}$, $T_{0}$, $\gamma$ (with $1.0 \le \gamma \le
      1.6$), and cosmological parameters, while the \citet{fg2008c}
      errors reflect only uncertainties in $\tau_{\rm eff}$ and
      $T_{0}$.  We find roughly a factor of two higher values of
      \G\ compared to \citet{fg2008c} due to the fact that we use
      measured temperatures that are significantly lower, and because
      our values are derived from artificial spectra drawn from
      hydrodynamical simulations rather than purely from a density
      PDF.  See text for details.  The evolution of \G\ at $z \ge 5$
      is discussed in Section~\ref{sec:discussion}.}
   \label{fig:G12_with_lit}
   \end{center}
   \end{minipage}
\end{figure*}

We compare our results to previous measurements of \G\ based on the
effective \lya\ optical depth form \citet{bolton2005},
\citet{bolton2007}, and \citet{fg2008c} in
Figure~\ref{fig:G12_with_lit} .  To facilitate a direct comparison, in
each case we plot our nominal value of \G\ for the fiducial value of
$\gamma$ adopted in those works.  The literature results have also
been adjusted for cosmology.

\begin{table}
   \renewcommand{\arraystretch}{1.1}
   \caption{Covariance matrix for the statistical errors in $\log{\Gamma}$.  Values have been multiplied by 100.}
   \vspace{-6pt}
   \label{tab:Gamma_covar}
   \begin{center}
   \begin{tabular*}{0.47\textwidth} {@{\extracolsep{\fill}}c|ccccccc}
   \hline
   z     &   2.40  &   2.80  &   3.20  &   3.60  &   4.00  &   4.40  &   4.75  \\
   \hline
   2.40  &   0.218  &  0.142  &  0.112  &  0.091  &  0.079  &  0.073  &  0.069 \\
   2.80  &          &  0.134  &  0.097  &  0.074  &  0.068  &  0.064  &  0.057 \\
   3.20  &          &         &  0.127  &  0.085  &  0.063  &  0.064  &  0.064 \\
   3.60  &          &         &         &  0.141  &  0.086  &  0.058  &  0.062 \\
   4.00  &          &         &         &         &  0.181  &  0.107  &  0.063 \\
   4.40  &          &         &         &         &         &  0.267  &  0.164 \\
   4.75  &          &         &         &         &         &         &  0.488 \\
   \hline
   \end{tabular*}
   \end{center}
\end{table}

Our results are consistent with those of \citet{bolton2005} over $3 <
z < 4$.  This is not surprising since the \taueffa\ and temperature
values adopted in that work were similar to the ones used here, though
with larger errors.  A similar technique using artificial spectra was
also used to determine the ionization rate.  A possible break between
our value of \G\ at $z=4.75$ and that of \citet{bolton2007} at $z=5$
may be present, but we defer discussion of the evolution of \G\ at $z
> 5$ to Section~\ref{sec:discussion}.

We are systematically higher by roughly a factor of two than the
results of \citet{fg2008c}.  Much of this difference can be attributed
to differences in the gas temperatures.  \citet{fg2008c} adopt $T_{0}$
values from \citet{zaldarriaga2001}, which, for their adopted value of
$\gamma = 1.6$, are roughly a factor of two larger than the values
measured by \citet{becker2011a}.  The higher temperatures will produce
values of \G\ that are $\sim$0.2 dex lower.  Part of the remaining
difference may relate to the different methods used to compute \G.
\citet{fg2008c} derived their \G\ values by directly integrating
\lya\ optical depths over a density distribution function in order to
predict the mean transmitted flux.  This neglects the effects of
peculiar velocities and thermal broadening, which we incorporate by
using mock spectra drawn from numerical simulations.  Peculiar
velocities tend to consolidate optical depth in redshift space due to
the infall of gas towards filaments, thus requiring a lower ionization
rate to achieve a given mean flux.  In tests where we ignored peculiar
velocities we obtained a higher \G\ by $\sim$3 (30) per cent at $z
\simeq 2$ (4).  Thermal broadening, in contrast, tends to increase the
amount of absorption by lines of high optical depth.  When we ignored
thermal broadening, our results for \G\ decreased by $\sim$38 (26) per
cent at $z \simeq 2$ (4).  The net result of ignoring both peculiar
velocities and thermal broadening was to decrease \G\ by $\sim$30 per
cent at $z \simeq 2$ while leaving it essentially unchanged at $z
\simeq 4$.  This may help to explain the somewhat larger difference
between our results and those of \citet{fg2008c} at $z \sim 2.5$.  We
note that the \taueffa\ values used by both works are generally
consistent, and do not contribute greatly to the difference in
\G\ where the results overlap.

\section{Ionizing Emissivity}\label{sec:emissivity}

\subsection{Method}\label{sec:emissivity_method}

In this section we turn to inferring the metagalactic ionizing
emissivity from our photoionization rate measurements.  The hydrogen
ionization rate inferred from the \lya\ forest opacity reflects the
intensity of the UVB after processing by the IGM.  In order to use
this to infer the net rate at which ionizing photons are being emitted
by all sources, radiative transfer effects must be taken into account.
The ionization rate is related to the mean specific intensity,
$J(\nu)$, by equation~(\ref{eq:Gamma}).  For our calculations we adopt
$\sigma_{\rm H\, I}(\nu) \propto \nu^{-2.75}$, which is closer to the
true frequency dependence \citep[e.g.,][]{osterbrock2006} than the
commonly-used $\nu^{-3}$ approximation, at least over the emitted
frequency range that dominates the \hi\ ionization rate ($1\,{\rm Ryd}
\le h\nu \lesssim 2\,{\rm Ryd}$).  This choice, however, has little
impact on our results.  The mean specific intensity at redshift
$z_{0}$ and frequency $\nu_{0}$, in turn, can be written as an
integral over the emissivity, $\epsilon(\nu,z)$, from all sources at
$z \ge z_{0}$ as \citep[e.g.,][]{hm1996}
\begin{equation}
J(\nu_{0},z_{0}) = \frac{1}{4\pi} \int_{z_{0}}^{\infty} dz \frac{dl}{dz} \frac{(1+z_{0})^3}{(1+z)^3} \epsilon(\nu,z) e^{-\tau_{\rm eff}(\nu_{0},z_{0},z)} \, ,
\label{eq:specific_intensity}
\end{equation}
where $\nu = \nu_{0}(1+z)/(1+z_{0})$, $dl/dz = c/[(1+z)H(z)]$ is the
proper line element, and $\tau_{\rm eff}(\nu_{0},z_{0},z)$ is the
effective optical depth for photons with frequency $\nu_{0}$ at
redshift $z_{0}$ that were emitted at redshift $z$.  We note that
equation~(\ref{eq:specific_intensity}) takes into account cosmological
radiative transfer effects, including the redshifting of ionizing
photons.

It is common to simplify equation~(\ref{eq:specific_intensity}) by
assuming that the mean free path to ionizing photons is sufficiently
short that the ionizing sources are essentially local.  In this case,
redshifting effects may be neglected and $J(\nu,z) \approx
(1/4\pi)\lambda_{\rm mfp}(\nu,z)\epsilon(\nu,z)$, where $\lambda_{\rm
  mfp}$ is the mean free path. As we demonstrate in
Appendix~\ref{app:em_calc}, however, this approximation can produce
considerable underestimates of the ionizing emissivity, up to a factor
of two at $z \sim 2$.  This occurs because the local source
approximation neglects the fact that at $z < 4$ a significant fraction
of the ionizing photons emitted will redshift beyond the Lyman limit
without being absorbed by the IGM.  This loss of ionizing photons must
be taken into account in order to calculate the total ionizing
emissivity from the hydrogen ionization rate.

\begin{figure*}
   \centering
   \begin{minipage}{\textwidth}
   \begin{center}
   \includegraphics[width=0.75\textwidth]{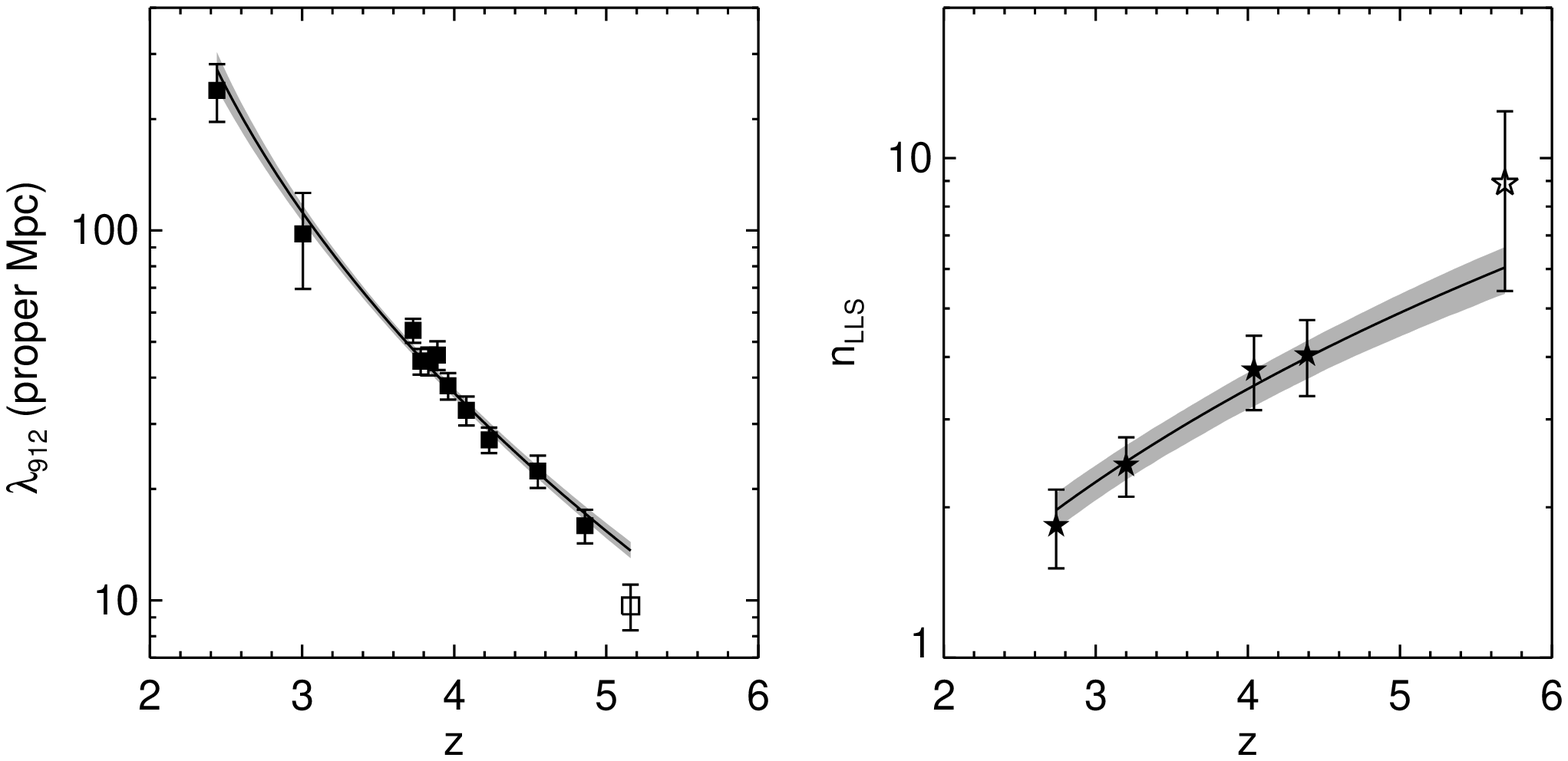}
   \vspace{-0.05in}
   \caption{Parameters describing the opacity of the IGM to ionizing
     photons.  The left panel shows the mean free path at 912~\AA,
     with data points taken from \citet{omeara2013} ($z=2.4$),
     \citet{fumagalli2013} ($z=3.0$), \citet{prochaska2009b}
     ($3.73 \ge z \ge 4.23$), and Worseck et al., in prep ($4.55 \ge z
     \ge 5.16$), with updates as compiled in Worseck et al..  The
     right-hand panel shows the number density of Lyman limit systems,
     with data from \citet{songaila2010}.  The solid lines in each
     panel show our fit to these quantities assuming that the
     \hi\ column density distribution has the form $f(N_{\rm H\,I},z)
     \propto N_{\rm H\,I}^{-\beta_{\rm N}}(1+z)^{\beta_{z}}$.  Shaded
     regions show the 1$\sigma$ uncertainty in the fit.  Data with
     empty symbols were not included in the fit, as they lie outside
     the redshift range covered by our analysis.  For the sake of
     comparison, however, we plot the fits extrapolated to these
     redshifts.}
   \label{fig:cddf}
   \end{center}
   \end{minipage}
\end{figure*}

For Poisson-distributed absorbers with a column density distribution
$f(N_{\rm H,I},z) = \partial^{2}\mathcal{N}/\partial N_{\rm H,I}
\partial z$, $\tau_{\rm eff}(\nu_{0},z_{0},z)$ can be calculated as
\citep{paresce1980}
\begin{equation}
\tau_{\rm eff}(\nu_{0},z_{0},z) = \int_{z_{0}}^{z} dz' \int_{0}^{\infty} dN_{\rm H\,I} f(N_{\rm H\,I},z') (1 - e^{-\tau_{\nu}}) \, ,
\label{eq:taueff_ion} 
\end{equation}
where $\tau_{\nu} = N_{\rm H\,I}\sigma_{\nu}$.  It has been shown that
$f(N_{\rm H,I},z)$ has a complex shape over column densities that
provide most of the optical depth to ionizing photons ($10^{15}~{\rm
  cm^{-2}} \lesssim N_{\rm H\,I} \lesssim 10^{20}~{\rm cm^{-2}}$)
\citep[e.g.,][]{omeara2007,prochaska2010}.  Ideally one would
integrate equation~(\ref{eq:taueff_ion}) over a detailed fit to the
column density distribution \citep[e.g.,][]{hm2012}; however, large
uncertainties exist in the amplitude of $f(N_{\rm H,I},z)$ at column
densities within the range of interest, up to an order of magnitude at
$N_{\rm H\,I} = 10^{17}\,{\rm cm^{-2}}$
\citep{prochaska2010,omeara2013}.  Rather than attempt to integrate
over the detailed column density distribution, therefore, we follow
the common convention of approximating $f(N_{\rm H\,I},z)$ as a power
law of the form
\begin{equation}
f(N_{\rm H\,I},z) = \frac{A}{N_{\rm LL}} \left(\frac{N_{\rm H\,I}}{N_{\rm LL}}\right)^{-\beta_{N}} \left(\frac{1+z}{4.5}\right)^{\beta_{z}} \,
\label{eq:cddf}
\end{equation}
where $N_{\rm LL} = 10^{17.2}~{\rm cm^{-2}}$ is the Lyman limit column
density.  Although this is a rough approximation, we have the
advantage that recent works analyzing composite quasar spectra have
provided direct measurements of the integrated opacity at $h\nu_{0} =
1$ Ryd, independent of the shape of $f(N_{\rm H\,I},z)$.  With these
measurements as an anchor point, our main requirement is to model
$f(N_{\rm H\,I},z)$ with sufficient accuracy to determine the opacity
at higher energies.  This depends mainly on the relative contribution
to $\tau_{\rm eff}(\nu_{0},z_{0},z)$ from optically thin and thick
systems.  If the ionizing opacity is mainly due to optically thin
systems, then the mean free path as a function of frequency will scale
inversely with the ionization cross section, $\lambda_{{\rm
    mfp},\nu}^{\rm thin} \propto \sigma_{\nu}^{-1} \propto
\nu^{2.75}$.  If instead the opacity comes from optically thick
`bricks', then the mean free path will be independent of frequency.  A
power law approximation to the column density distribution provides a
convenient means to quantify this balance, at least to first order, as
well as to simplify the calculation of $\tau_{\rm
  eff}(\nu_{0},z_{0},z)$.

We determine the parameters for $f(N_{\rm H\,I},z)$ in
equation~(\ref{eq:cddf}) by combining direct measurements of the mean
free path for 1 Ryd photons, \mfp, with measurements of the number
density of Lyman limit systems ($N_{\rm H\,I} \ge 10^{17.2}~{\rm
  cm^{-2}}$).  The \mfp\ values were determined directly from the flux
profiles of composite quasar spectra by \citet{prochaska2009b},
\citet{omeara2013}, \citet{fumagalli2013}, and Worseck et al. (in
prep).  We note that $\lambda(z)$ is defined here as a function of the
emitted redshift, in the sense that a packet of photons emitted at
redshift $z$ with frequency $\nu = \nu_{0}(1+z)/(1+z_{0})$ travels a
distance $\lambda$ to redshift $z_{0}$, by which point their frequency
is $\nu_{0}$ and the number of photons is diminished, on average, by a
factor of $e$.  Thus, \mfp\ refers to the mean free path for photons
emitted with energy greater than 1 Ryd.  Values for the number density
of Lyman limit systems, $n_{\rm LLS}$, were taken from
\citet{songaila2010}, who also included measurements compiled by
\citet{peroux2003}.  We note that other works have used $n_{\rm LLS}$
to infer \mfp, which requires assuming a slope for $f(N_{\rm H\,I})$
near the Lyman limit \citep[e.g.,][]{madau1999,kuhlen2012}.  Here,
however, we are combining independent measurements of \mfp\ and
$n_{\rm LLS}$ to constrain $\beta_{N}$.  We emphasize again that this
is really an effective slope for $f(N_{\rm H\,I},z)$ intended only to
facilitate the calculation of the ionizing opacity as a function of
frequency and redshift.  It will therefore not reflect the full
details of the \hi\ column density distribution.

The parameters $A$, $\beta_{N}$, and $\beta_{z}$ are determined by
simultaneously fitting the \mfp\ and $n_{\rm LLS}$ values as a
function of redshift.  The number density of Lyman limit systems is
obtained by directly integrating $f(N_{\rm H\,I},z)$ over column
density to obtain $n_{\rm LLS}(z)$.  We use
equation~(\ref{eq:taueff_ion}) to determine $z_{0}$ at which $\tau_{\rm
  eff}(\nu_{912},z_{0},z) = 1$, and then convert this into a mean free
path at the Lyman limit by computing the proper distance between $z$
and $z_{0}$.  We integrate equation~(\ref{eq:taueff_ion}) over a
finite range in \hi\ column density with nominal limits of $[N_{\rm
    H\,I}^{\rm min},N_{\rm H\,I}^{\rm max}] = [10^{15},10^{21}]~{\rm
  cm^{-2}}$.  These limits are meant to span the range in $N_{\rm H\,
  I}$ that dominates the optical depth to ionizing photons
\citep[e.g.,][and references therein]{hm2012}, such that our power-law
approximation for $f(N_{\rm H\,I},z)$ most closely matches the true
shape of the column density distribution over this range.  We found
our results for $f(N_{\rm H\,I},z)$ and the emissivity to be
insensitive to the lower limit of integration for $N_{\rm H\,I}^{\rm
  min} \le 10^{16}~{\rm cm^{-2}}$.  Increasing the upper limit will
tend to flatten $f(N_{\rm H})$ and decrease A.  This ultimately has
only a minor impact on the calculated emissivity, however, as
quantified below.

Our best-fitting parameters for $f(N_{\rm H\,I},z)$ are $[A,\beta_{\rm
    N},\beta_{z}] = [0.93 \pm 0.08,1.33 \pm 0.05, 1.92 \pm 0.15]$,
where the errors in $A$ and $\beta_{N}$ are highly correlated.  The
accompanying fits to $\lambda_{912}(z)$ and $n_{\rm LLS}(z)$ are
plotted in Figure~\ref{fig:cddf}.  This fit also produces a number
density of LLSs with $\tau > 2$ that is consistent at the
$<$1.5$\sigma$ level with measurements over $2 \lesssim z \lesssim 4$
\citep{prochaska2010,omeara2013,fumagalli2013}.  Uncertainties in the
parameters were determined by perturbing the data points according to
their errors and re-calculating the fit.  The corresponding errors in
our fits to $\lambda_{912}(z)$ and $n_{\rm LLS}(z)$ are shown as
shaded regions in Figure~\ref{fig:cddf}.\footnote{We note that despite
  the fact that we are consistent with the $n_{\rm LLS}(z)$ values of
  \citet{songaila2010}, by design, and our best-fit value of
  $\beta_{\rm N}$ is similar to the one used by those authors, the
  values of \mfp\ adopted here are considerably larger than those
  derived by them and adopted by \citet{kuhlen2012} at $z \le 3.4$.
  For example, the mean free path we determine from our $f(N_{\rm
    H\,I},z)$ fit at $z = 2.6$ is $\sim$50 per cent larger than the
  \citet{songaila2010} value.  This is primarily because the
  \citet{songaila2010} values are calculated using an expression that
  assumes $z_{0} \simeq z$ (their equation (7); see also
  \citet{madau1999}, \citet{fg2008c} equation (22)).  This
  underestimated the true mean free path as defined above,
  particularly at $z < 4$.}
   
We note that our results are not highly sensitive to exact value of
$\beta_{N}$, provided that we fit the \mfp\ values.  For example, if
we fix $\beta_{N}$ to be 1.5 \citep[e.g.,][and references
  therein]{madau1999}, and refit for A and $\beta_{z}$, our estimates
of the emissivity would decrease by less than 12 per cent versus the
nominal values.  Doing so would underestimate the number density of
LLSs.  To be conservative, however, we adopt an additional systematic
uncertainty in the emissivity related to fitting $f(N_{\rm H\,I},z)$
of $\pm 0.05$ dex.

In order to address whether approximating $f(N_{\rm H\,I},z)$ as a
single power law may bias our emissivity estimates, we performed the
following test.  We first generated an artificial data set by
calculating the predicted values for $\lambda_{912}(z)$ and $n_{\rm
  LLS}(z)$ over $2.4 < z < 5$ from the piece-wise fit to $f(N_{\rm
  H\,I},z)$ adopted by \citet{hm2012}.  We then fit a single power law
$f(N_{\rm H\,I},z)$ to these values following the procedure described
above, finding $[A,\beta_{\rm N},\beta_{z}] = [1.05,1.52,1.91]$.  Next
we calculated the emissivity using the \citet{hm2012} piece-wise fit
directly and using the single power law approximation.  For a fixed
$\Gamma(z)$, the single power law produced emissivity results that
were 5-19 per cent lower than those obtained using the piece-wise
$f(N_{\rm H\,I},z)$.  The bias was found to increase with increasing
redshift and for harder ionizing spectra (see below).  These results
suggest that for the narrow purpose of estimating the emissivity near
1 Ryd, a single power law model for $f(N_{\rm H\,I},z)$ is a
reasonable approximation provided that it reproduces the correct
\mfp\ and $n_{\rm LLS}$ values.  The precise level of bias will depend
on the details of the true \hi\ column density distribution.  Taking
these results as a guide, however, we apply a positive 10 percent
correction to our raw emissivity values, and increase the systematic
uncertainty related to fitting $f(N_{\rm H\,I},z)$ by a further $\pm
0.05$ dex.

We follow the common convention of modeling the spectral shape of the
specific emissivity as a power law of the form
\begin{equation}
\epsilon(\nu,z) = \epsilon_{912}(z) \left( \frac{\nu}{\nu_{912}} \right)^{-\alpha} \, ,
\label{eq:specific_emissivity}
\end{equation}
where $\epsilon_{912}$ is the specific emissivity at the Lyman limit
and $\alpha$ is the spectral index blueward of the Lyman limit.  The
integrated emissivity of ionizing photons, $\dot{N}_{\rm ion}$, is
then given by
\begin{equation}
\dot{N}_{\rm ion}(z) = \int_{\nu_{912}}^{\infty} d\nu \frac{\epsilon(\nu,z)}{h\nu} = \frac{\epsilon_{912}}{h\alpha} \, .
\label{eq:Nion}
\end{equation}

The most appropriate value for $\alpha$ is a subject of debate.  For
AGN, a value of $\alpha = 1.6$ \citep{telfer2002} is commonly used,
whereas for galaxies the adopted values of $\alpha$ range between 1
and 3 \citep[e.g.,][]{bolton2007,ouchi2009,kuhlen2012}.  For realistic
galaxy spectra, however, a single power law is probably an
oversimplification, even between the \hi\ and \heii\ ionizing edges.
For example, a {\sc starburst99} model \citep{leitherer1999} with
$t_{\rm age} = 500$~Myr\footnote{The figures given here do not depend
  sensitively on age, since the spectral shape blueward of 912~\AA\ is
  dominated by short-lived, massive stars.}, continuous star
formation, Salpeter IMF, and metallicity $Z = (1/5)Z_{\odot}$ has a
slope of $\alpha \simeq 1.3$ between 1 and 1.8~Ryd.  This energy range
is responsible for most of the \hi\ photoionization events due to the
steep frequency dependence of the ionizing cross section, and so this
value of $\alpha$ is appropriate for calculating \eps.  At higher
energies, however, the spectrum softens and essentially goes to zero
at $E > 4$~Ryd.  Integrating the model spectrum over all frequencies
$\nu \ge \nu_{912}$ gives the same total number of photons as
integrating over a pure power-law spectrum with $\alpha =
1.8$.\footnote{This {\sc starburst99} was previously used to motivate
  a model in which $\alpha = 3.0$
  \citep[e.g.,][]{bolton2007,ouchi2009,nestor2013}.  This value
  reflects the ratio of luminosities at the \hi\ and \heii\ ionizing
  edges; however, between these limits the model contains considerably
  more flux than would an $L_{\nu} \propto \nu^{-3}$ power law with
  the same luminosity just below 912~\AA.}  For a {\sc starburst99}
model with $Z = (1/20)Z_{\odot}$, the slope near 1~Ryd is $\sim$1.1,
while the ``effective'' slope integrating over all frequencies is
$\sim$1.7.  For $Z = Z_{\odot}$ these values become roughly 1.5 and
2.0, respectively.  Recent models using the {\sc bpass} spectral
library \citep{eldridge2012} and including massive binaries give
similar results at low metallicities, though the ionizing spectra are
softer at solar metallicity.\footnote{We note, however, that the
  spectral break between 1500~\AA\ and 912~\AA, $L_{1500}/L_{912}$, is
  significantly smaller in the {\sc bpass} models for all
  metallicities.}  These examples illustrates that care must be taken
when converting \G, which is dominated by photons close to 1~Ryd, to a
specific ionizing emissivity near the Lyman limit and/or a total
ionizing emissivity.  For simplicity, however, we will mainly treat
the emitted ionizing spectrum as a single power law at $h\nu \ge 1$
Ryd.  We adopt a nominal value of $\alpha = 2.0$, but also calculate
\eps\ and \Nion\ for $\alpha = 1.0$ and 3.0.  In addition, we perform
calculations for the {\sc starburst99} model galaxy spectrum with $Z =
(1/5)Z_{\odot}$ mentioned above.

 \begin{table*}
   \renewcommand{\arraystretch}{1.3}
   \caption{Results and error budget for $\log{\epsilon_{912}}$ and
     $\log{\dot{N}_{\rm ion}}$.  Units for \eps\ are $10^{23}\,{\rm
       erg\,s^{-1}\,Hz^{-1}\,Mpc^{-3}}$.  Units for \Nion\ are
     $10^{51}\,{\rm photons\,s^{-1}\,Mpc^{-3}}$.  The nominal values
     are for $\gamma = 1.4$, $\alpha = 2.0$.  Statistical errors
     reflect the diagonal terms of the covariance matrix only.
     Systematic errors in $f(N_{\rm H\,I},z)$ include errors arising
     from changes to $[N_{\rm H\,I}^{\rm min},N_{\rm H\,I}^{\rm max}]$,
     possible systematic changes to $\beta_{N}$, and the approximation
     of $f(N_{\rm H\,I},z)$ as a single power law (see text).}
   \vspace{-6pt}
   \label{tab:emissivity_errors}
   \begin{center}
   \begin{tabular*}{\textwidth} {@{\extracolsep{\fill}}l*{4}{c}@{\hskip 0.6cm}*{4}{c}}
   \hline
      &  \multicolumn{4}{c}{$\log{\epsilon_{912}}$}   & \multicolumn{4}{c}{$\log{\dot{N}}$}  \\
   z                              	&        2.40  &        3.20  &        4.00  &        4.75  &        2.40  &        3.20  &        4.00  &        4.75  \\	  
   \hline
   Nominal value			&       2.076  &       1.911  &       1.983  &       2.108  &    $-$0.046  &    $-$0.211  &    $-$0.139  &    $-$0.014  \\										    
   \hline				                                                                                                                                                                                                          
   $\tau_{\rm eff}$ errors only		&  $\pm$0.048  &  $\pm$0.021  &  $\pm$0.015  &  $\pm$0.018  &  $\pm$0.048  &  $\pm$0.021  &  $\pm$0.015  &  $\pm$0.018  \\										    
   $T(\Delta)$ errors only        	&  $\pm$0.028  &  $\pm$0.032  &  $\pm$0.040  &  $\pm$0.040  &  $\pm$0.028  &  $\pm$0.032  &  $\pm$0.040  &  $\pm$0.040  \\										    
   $f(N_{\rm HI},z)$ stat. errors only  &  $\pm$0.011  &  $\pm$0.012  &  $\pm$0.014  &  $\pm$0.019  &  $\pm$0.011  &  $\pm$0.012  &  $\pm$0.014  &  $\pm$0.019  \\										    
   Cosmology errors only          	&  $\pm$0.017  &  $\pm$0.017  &  $\pm$0.019  &  $\pm$0.021  &  $\pm$0.017  &  $\pm$0.017  &  $\pm$0.019  &  $\pm$0.021  \\										    
   Total statistical error        	&  $\pm$0.059  &  $\pm$0.044  &  $\pm$0.048  &  $\pm$0.053  &  $\pm$0.059  &  $\pm$0.044  &  $\pm$0.048  &  $\pm$0.053  \\										    
   $\gamma = 1.6$                 	&    $+$0.084  &    $+$0.089  &    $+$0.074  &    $+$0.065  &    $+$0.084  &    $+$0.089  &    $+$0.074  &    $+$0.065  \\										    
   $\gamma = 1.2$                 	&    $-$0.084  &    $-$0.089  &    $-$0.075  &    $-$0.065  &    $-$0.084  &    $-$0.089  &    $-$0.075  &    $-$0.065  \\										    
   $\alpha = 3.0$                 	&    $+$0.125  &    $+$0.123  &    $+$0.119  &    $+$0.113  &    $-$0.052  &    $-$0.053  &    $-$0.057  &    $-$0.063  \\										    
   $\alpha = 1.0$                 	&    $-$0.153  &    $-$0.155  &    $-$0.151  &    $-$0.145  &    $+$0.148  &    $+$0.146  &    $+$0.150  &    $+$0.156  \\										    
   Jeans smoothing                	&  $_{-0.022}^{+0.000}$  &  $_{-0.006}^{+0.010}$  &  $_{-0.003}^{+0.022}$  &  $_{-0.010}^{+0.022}$  &  $_{-0.022}^{+0.000}$  &  $_{-0.006}^{+0.010}$  &  $_{-0.003}^{+0.022}$  &  $_{-0.010}^{+0.022}$  \\
   $f(N_{\rm H\,I},z)$ systematics  	&  $_{-0.106}^{+0.103}$  &  $_{-0.111}^{+0.106}$  &  $_{-0.113}^{+0.107}$  &  $_{-0.114}^{+0.108}$  &  $_{-0.106}^{+0.103}$  &  $_{-0.111}^{+0.106}$  &  $_{-0.113}^{+0.107}$  &  $_{-0.114}^{+0.108}$  \\ 
   Recombination rad. systematics        &  $\pm$0.050  &  $\pm$0.050  &  $\pm$0.050  &  $\pm$0.050  &  $\pm$0.050  &  $\pm$0.050  &  $\pm$0.050  &  $\pm$0.050  \\
   Total systematic error         	&  $_{-0.415}^{+0.362}$  &  $_{-0.411}^{+0.378}$  &  $_{-0.392}^{+0.372}$  &  $_{-0.383}^{+0.358}$  &  $_{-0.313}^{+0.385}$  &  $_{-0.308}^{+0.401}$  &  $_{-0.298}^{+0.403}$  &  $_{-0.301}^{+0.401}$  \\
   Total error                    	&  $_{-0.474}^{+0.421}$  &  $_{-0.455}^{+0.422}$  &  $_{-0.440}^{+0.420}$  &  $_{-0.437}^{+0.412}$  &  $_{-0.372}^{+0.444}$  &  $_{-0.352}^{+0.445}$  &  $_{-0.346}^{+0.451}$  &  $_{-0.355}^{+0.454}$  \\
   \hline
   \end{tabular*}
   \end{center}
\end{table*}

We combine equations~(\ref{eq:Gamma})-(\ref{eq:Nion}) in order to
calculate \eps\ and \Nion\ from our \G\ values and the ionizing
opacity measurements described above.  We use an analytic solution for
$\tau_{\rm eff}(\nu_{912},z_{0},z)$, which is possible for a power-law
$f(N_{\rm H\,I},z)$.\footnote{See \citet{fg2008c} for the case in
  which $[N_{\rm H\,I}^{\rm min},N_{\rm H\,I}^{\rm max}] =
  [0,\infty]$.}  The remainder of our calculations are performed
numerically.  Since we are not using the local-source approximation,
$\Gamma(z_{0})$ will depend on $\epsilon_{912}(z)$ at $z \ge z_{0}$,
and so $\epsilon_{912}(z)$ must be modeled as a continuous function in
redshift.  We therefore parametrize $\epsilon_{912}(z)$ at four
discreet redshifts between $z = 2.4$ and 4.75, and assume that it
evolves linearly between these points.
 
Finally, we apply an approximate correction to the emissivity to
account for recombination radiation.  For a given emissivity from
galaxies and AGN, recombination radiation from the IGM will tend to
enhance the photoionization rate
\citep{hm1996,hm2012,fardal1998,fg2009}.  We estimate the fractional
increase in \G\ when including this effect using the analytic
expressions in \citet{fg2009} (their Appendix C).  For the $f(N_{\rm
  H\,I},z)$ parameters used here, and assuming a gas temperature of
10,000 K (20,000 K), we find $\Gamma^{\rm with\,rec}/\Gamma^{\rm no\,
  rec} \simeq 1.05-1.12$ ($1.08-1.15$) over $2.4 \le z \le 4.75$,
where the effect increases with redshift.  This is similar to, though
slightly lower than, the amplitude found by \citet{fg2009} for their
fiducial set of parameters.  As explored by those authors, the precise
impact of recombination radiation will depend on the details of the
\hi\ column density distribution and the temperature structure of the
IGM, which are only approximately modeled here.  As a first-order
correction, therefore, we simply decrease our emissivity results by 10
per cent (offsetting the above correction related to $f(N_{\rm
  H\,I},z)$), but include an additional $\pm$0.05 dex systematic
uncertainty.

\subsection{Emissivity results}\label{sec:emissivity_results} 
 
\begin{table}
   \renewcommand{\arraystretch}{1.1}
   \caption{Covariance matrix for the statistical errors in
     $\log{\epsilon_{912}}$ and $\log{\dot{N}_{\rm ion}}$.}
   \vspace{-6pt}
   \label{tab:emissivity_covar}
   \begin{center}
   \begin{tabular*}{0.35\textwidth} {@{\extracolsep{\fill}}c|cccc}
   \hline
   z     &  2.40  &  3.20  &  4.00  &  4.75  \\
   \hline
   2.40  &  3.46  &  0.74  &  1.18  &  0.49  \\
   3.20  &        &  1.95  &  0.03  &  0.69  \\
   4.00  &        &        &  2.28  &  0.19  \\
   4.75  &        &        &        &  2.84  \\
   \hline
   \end{tabular*}
   \end{center}
\end{table}

\begin{figure*}
   \centering
   \begin{minipage}{\textwidth}
   \begin{center}
   \includegraphics[width=0.70\textwidth]{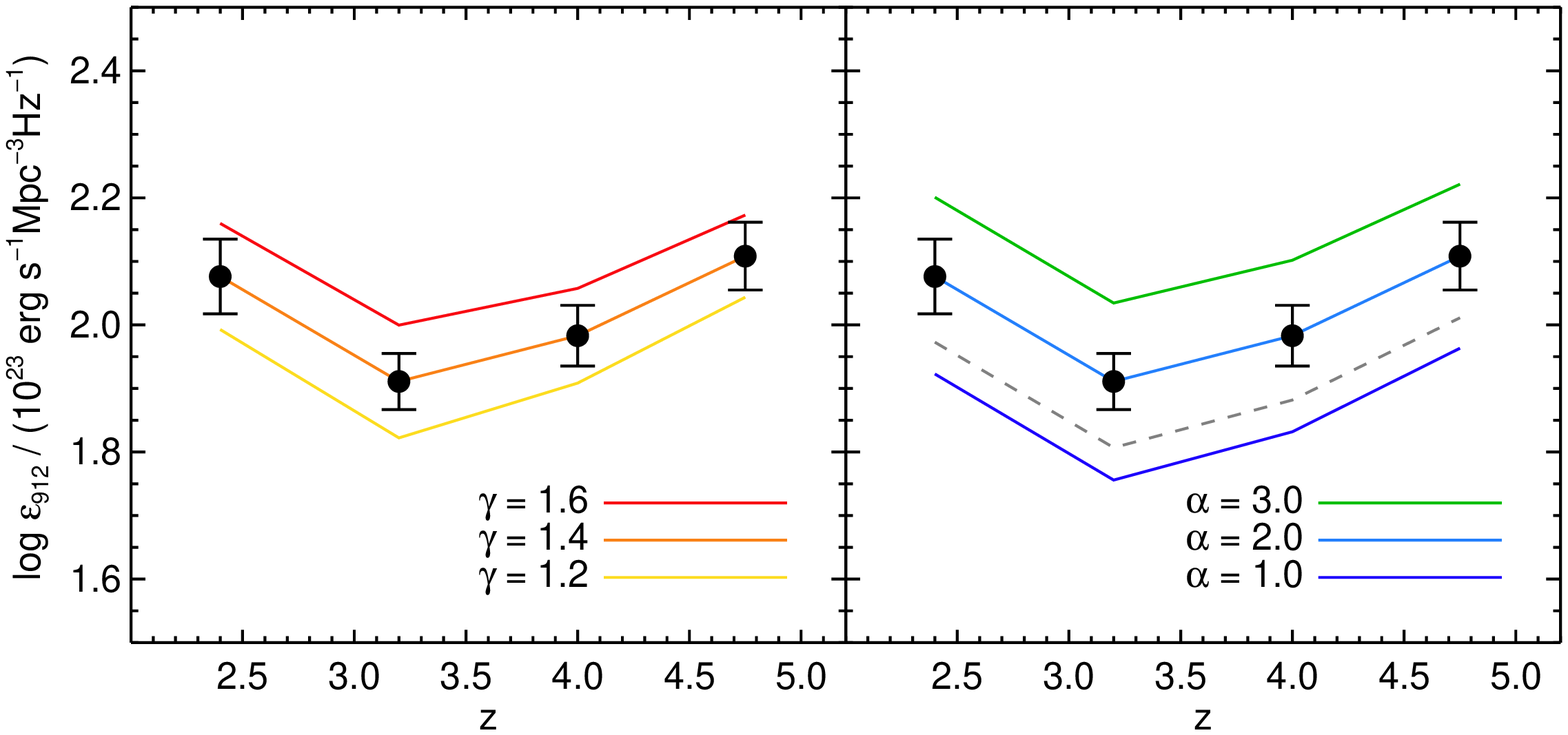}
   \vspace{-0.05in}
   \caption{The specific emissivity near the Lyman limit as a function
     of redshift.  Filled circles with 1$\sigma$ statistical error
     bars show \eps\ for our fiducial temperature-density parameter
     ($\gamma = 1.4$) and spectral index of the ionizing sources
     ($\alpha = 2.0$).  In the left-hand panel, solid lines give
     \eps\ for $\gamma = [1.6,1.4,1.2]$ (top to bottom).  In the
     right-hand panel, solid lines give \eps\ for $\alpha =
     [1.0,2.0,3.0]$ (bottom to top).  The dashed line gives
     approximate values for a {\sc starburst99} model with continuous
     star formation and $Z = (1/5)Z_{\odot}$ (see text).  Additional,
     smaller systematic uncertainties are summarized in
     Table~\ref{tab:emissivity_errors}.}
   \label{fig:eps23}
   \end{center}
   \end{minipage}
\end{figure*}

\begin{figure*}
   \centering
   \begin{minipage}{\textwidth}
   \begin{center}
   \includegraphics[width=0.70\textwidth]{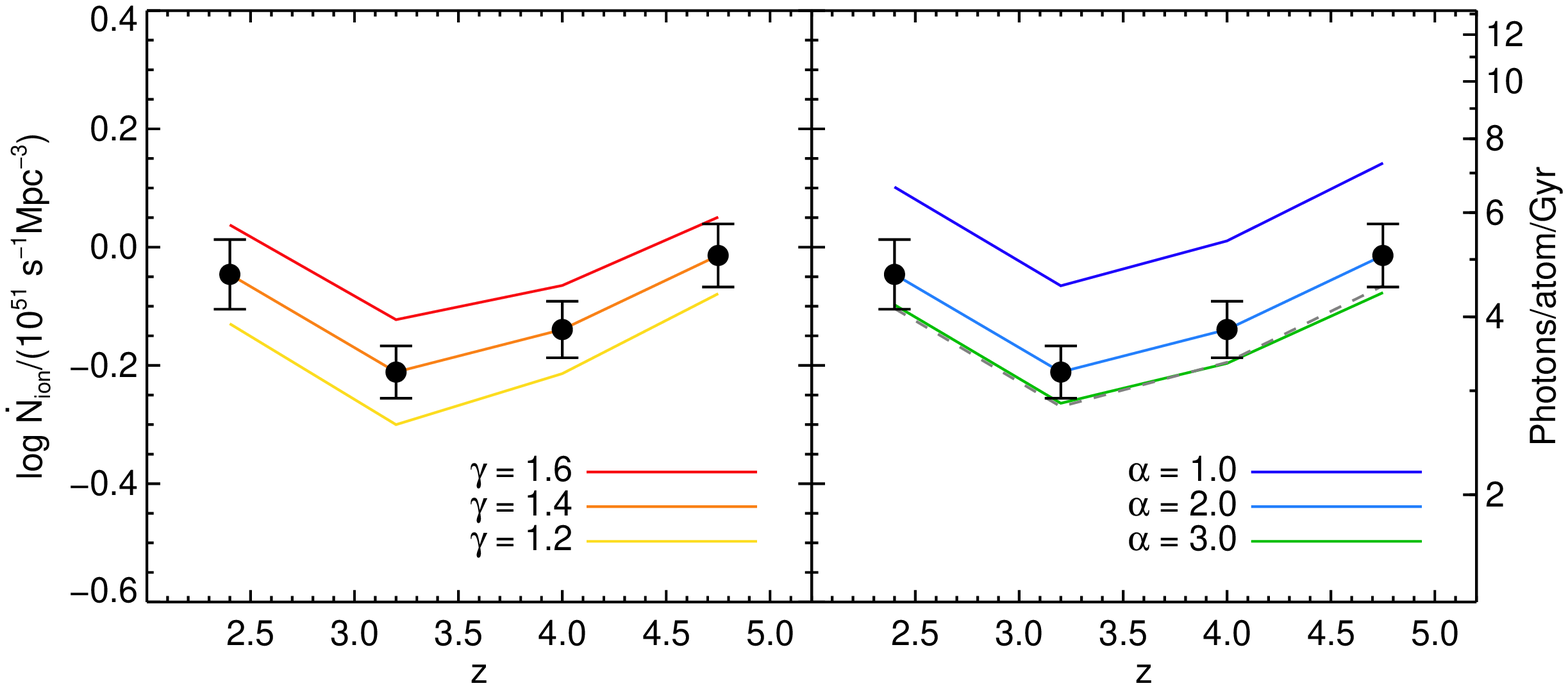}
   \vspace{-0.05in}
   \caption{The integrated emissivity of ionizing photons as a
     function of redshift.  Filled circles with 1$\sigma$ statistical
     error bars show \Nion\ for our fiducial temperature-density
     parameter ($\gamma = 1.4$) and spectral index of the ionizing
     sources ($\alpha = 2.0$).  In the left-hand panel, solid lines
     give \Nion\ for $\gamma = [1.6,1.4,1.2]$ (top to bottom).  In the
     right-hand panel, solid lines give \Nion\ for $\alpha =
     [1.0,2.0,3.0]$ (bottom to top).  The dashed line gives
     approximate values for a {\sc starburst99} model with continuous
     star formation and $Z = (1/5)Z_{\odot}$ (see text).  Additional,
     smaller systematic uncertainties are summarized in
     Table~\ref{tab:emissivity_errors}.}
   \label{fig:Nion}
   \end{center}
   \end{minipage}
\end{figure*}

\begin{figure*}
   \centering
   \begin{minipage}{\textwidth}
   \begin{center}
   \includegraphics[width=1.0\textwidth]{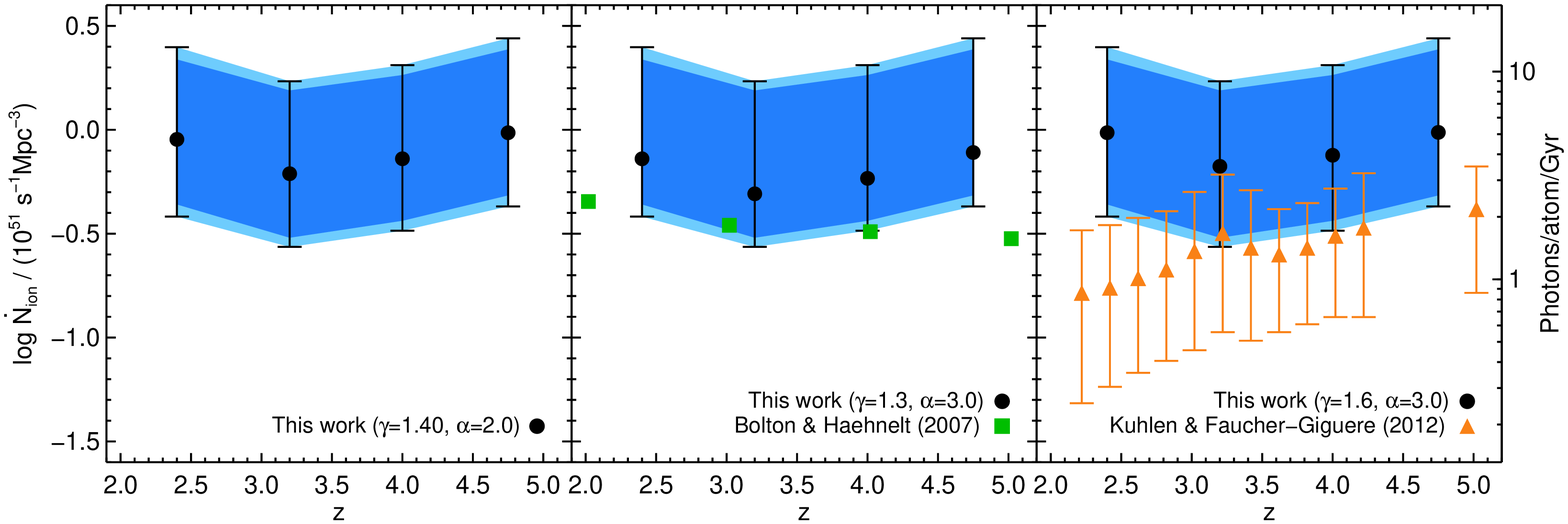}
   \vspace{-0.1in}
   \caption{Integrated emissivity of ionizing photons, \Nion, along
     with literature values over $2 < z < 5$.  In each panel the inner
     shaded band gives the total range of systematic uncertainty,
     while the outer shaded band gives the total statistical error.
     In the left-hand panel the filled circles show the nominal values
     of \Nion\ for our fiducial model with $\gamma = 1.4$ and $\alpha
     =2.0$.  In the middle and left-hand panels, the filled circles
     show \Nion\ for values of $\gamma$ and $\alpha$ corresponding to
     those adopted in previous works \citep{bolton2007,kuhlen2012}.
     The \citet{kuhlen2012} value at $z=5$ was calculated using
     \G\ from \citet{bolton2007}.  Literature values have been
     adjusted for cosmology and to reflect a $\sigma_{\nu} \propto
     \nu^{-2.75}$ scaling of the \hi\ ionization cross section.  The
     \citet{kuhlen2012} values have also been adjusted to give results
     integrated over all frequencies $\nu > \nu_{912}$.  Differences
     between our results and those of previous works are related mainly to
     the combined differences in \G\ and the ionizing opacity, as well
     as to the fact that we include radiative transfer effects when
     computing \Nion.  See text for details.}
   \label{fig:Nion_with_lit}
   \end{center}
   \end{minipage}
\end{figure*}

Our results for \eps\ and \Nion\ are presented in
Figures~\ref{fig:eps23} and \ref{fig:Nion}, respectively, with the
error budget summarized in Tables~\ref{tab:emissivity_errors} and
\ref{tab:emissivity_covar}.  The statistical errors in the emissivity
reflect the statical errors in \G\ (from \taueffa\ and
$T(\bar{\Delta})$), \mfp, and $n_{\rm LLS}$, as well in the cosmology.
Systematic errors from the slope of the temperature-density relation
and spectral slope are shown separately.  We also list smaller
systematic errors related to Jeans smoothing, 
$f(N_{\rm H\,I},z)$ fitting, and recombination radiation (see above).
Uncertainties in $\gamma$ produce systematic uncertainties in
\eps\ and \Nion\ that are comparable to the 1$\sigma$ statistical
errors.  Increasing $\alpha$ from 2.0 to 3.0 increases \eps\ by
0.11-0.13 dex, while decreasing $\alpha$ from 2.0 to 1.0 decreases
\eps\ by 0.13-0.16 dex.  Conversely, increasing $\alpha$ from 2.0 to
3.0 decrease \Nion\ by 0.05-0.07 dex, while decreasing $\alpha$ from
2.0 to 1.0 increases \Nion\ by 0.15-0.17 dex.  This occurs because a
harder emitted spectrum means that a larger proportion of ionizing
photons are emitted at frequencies where the ionization cross-section
is small, thus making a smaller contribution to \G.

We include estimates for \eps\ and \Nion\ for a {\sc starburst99}
model with continuous star formation and $Z = (1/5)Z_{\odot}$ in the
right-hand panels of Figures~\ref{fig:eps23} and \ref{fig:Nion}.  Note
that, in this case, adopting a more complex ionizing spectrum
decreases both \eps\ and \Nion.  This happens because the spectrum has
a relatively shallow slope near the Lyman limit, allowing a lower
\eps, but then steepens at higher energies, decreasing the total
\Nion.

Although \eps\ and \Nion\ may exhibit some evolution with redshift
within the statistical errors alone, the degree to which these
quantities truly change with redshift largely depends on how the
temperature-density relation slope, $\gamma$, and spectral shape of
the emissivity, $\alpha$, evolve.  Here were note, however, that the
total emissivity at $z = 4.75$ is quite substantial, with
$\dot{N}_{\rm ion} \sim 5$ ionizing photons per atom per
gigayear for our fiducial values of $\gamma$ and $\alpha$, or for the
example galaxy model spectrum.  \Nion\ may potentially be as high as
$\sim$14~${\rm photons/atom/Gyr}$ at $z=4.75$, but this is only for
$\alpha=1.0$, with no softening or break towards higher energies, and
a maximally steep temperature-density relation ($\gamma = 1.6$).

We compare our results to previous estimates of \Nion\ in
Figure~\ref{fig:Nion_with_lit}.  To facilitate a direct comparison, we
present our results for the same set of $\gamma$ and $\alpha$ values
used in the previous works\footnote{Values in \citet{kuhlen2012} Table
  2 are for $\alpha = 3.0$, although a fiducial values of 1.0 is
  stated in the text.}.  Minor corrections have been applied to the
literature results to reflect our adopted $\nu^{-2.75}$ frequency
dependence for the ionizing cross-section.  Corrections for cosmology
have also been applied, although we have not attempted to propagate
these through to the \citet{bolton2007} estimates of the mean free
path, which are based on modeling self-shielded gas in a
hydrodynamical simulation, and, as discussed by these authors, are
uncertain to within a factor of two.  We note that the errors on the
\citet{bolton2007} values of \Nion\ are probably too small due to the
degeneracy between \G\ and \mfp\ in their model; therefore, only their
nominal values for \Nion\ are shown in Figure~\ref{fig:Nion_with_lit}.
Finally, we have also made small adjustments to the \citet{kuhlen2012}
values such that \Nion\ is integrated over all frequencies $\nu \ge
\nu_{912}$, rather than only between the \hi\ and \heii\ ionizing
edges.

Our results are nominally $\sim$40 per cent higher than those of
\citet{bolton2007} at $z \simeq 3$.  This reflects the fact that we
obtain similar values for \G\ and \mfp\ at this redshift, while a
potential increase in the \citet{bolton2007} value of \Nion\ due to
the fact that they assume no modification of the ionizing spectrum due
to absorption by the IGM is more than offset by a decrease due to the
fact that they use a local-source approximation to calculate
\Nion\ (see Appendix~\ref{app:em_calc}).  At $z = 4$ their value falls
a factor of $\sim$1.8 below our result, mainly due to the fact that
they calculate a mean free path which is $\sim$60 per cent larger at
$z = 4$ than the one used here.  The factor of $\sim$2.5 difference at
$z \sim 5$ is due to their use of both a larger \mfp\ and a lower \G.

Our results generally overlap with those of \citet{kuhlen2012} within
the broad errors adopted by both works; however, our nominal values
for \Nion\ are a factor of two to six higher for the same values of
$\gamma$ and $\alpha$.  The factor of six discrepancy at $z=2.4$ is
partially explained by the fact that the \G\ value they adopt from
\citet{fg2008c} is a factor of $\sim$2.5 lower than our own at this
redshift.  As discussed above, this is due to the differences in the
gas temperatures and the fact that our results are based on artificial
spectra drawn from hydrodynamical simulations.  An additional factor
of two comes from the fact that they use a local source approximation
to compute the mean free path and \Nion\ (see
Appendix~\ref{app:em_calc}).  The remaining factor of $\sim$1.2
reflects small differences in the adopted shape of $f(N_{\rm H\,I},z)$
used to compute the mean free path \citep[see][]{songaila2010}, as
well as the fact that our \G\ values formally decrease from $z=2.4$ to
3.2, which amplifies the radiative transfer effect somewhat above the
case discussed in Appendix~\ref{app:em_calc}.

\section{The Sources of Ionizing Photons}\label{sec:sources}

Our estimates for the ionizing emissivity are based on the physical
conditions of the IGM and include the ionizing output from all
sources.  We now turn towards disentangling the contributions from AGN
and galaxies, and using the results to infer possible trends in the
ionizing efficiency of galaxies in the post-reionization era.

\begin{figure}
   \begin{center}
   \includegraphics[width=0.45\textwidth]{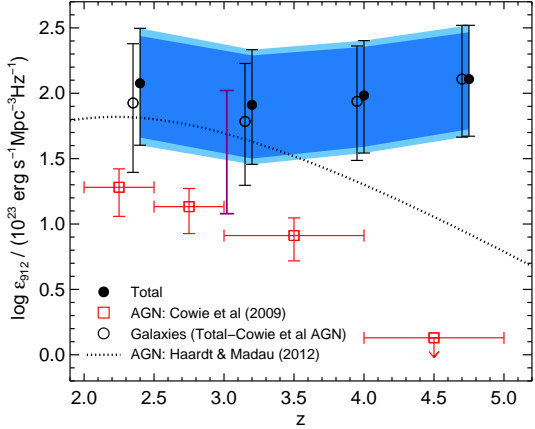}
   \vspace{-0.05in}
   \caption{The specific emissivity at 912~\AA.  Filled circles give
     our results for our fiducial parameters ($\gamma=1.4$,
     $\alpha=2.0$).  The inner shaded band gives the total range of
     systematic uncertainty, while the outer shaded band gives the
     total statistical error.  Estimates of the AGN emissivity from
     \citet{cowie2009} are shown as open squares, while the dotted
     line is the model AGN emissivity adopted by \citet{hm2012}.  The
     open circles give our results after subtracting the
     \citet{cowie2009} estimate of the AGN contribution.  The error
     bar at $z=3.0$ is an estimate of the contribution of galaxies to
     \eps\ based on direct measurements of escaping ionizing radiation
     from LBGs and LAEs (see text). }
   \label{fig:eps_with_AGN}
   \end{center}
\end{figure}

For the ionizing emissivity of AGN we adopt estimates made by
\citet{cowie2009}.  This work combined direct measurements of the
ionizing and near-UV luminosities of AGN at $z \sim 1$ with the
evolution of the near-UV luminosity density in an X-ray selected
sample of broad-line AGN over $0 < z < 5$.  We compare their results
for the specific emissivity from AGN at 912~\AA\ to our results for
the total specific emissivity from all sources in
Figure~\ref{fig:eps_with_AGN}.  The \citet{cowie2009} estimate of the
AGN contribution falls well below the total emissivity, and becomes an
increasingly small fraction towards higher redshifts.  We calculate
the contribution from galaxies, \epsgal, by subtracting the AGN
estimate from our total values, linearly interpolating the
\citet{cowie2009} AGN measurements onto our redshift bins and using a
Monte Carlo approach to propagate the errors.  Our estimates of the
galaxy emissivity are shown in Figure~\ref{fig:eps_with_AGN}.  The
galaxy and AGN contributions are potentially comparable, at least to
within the errors, at $z \simeq 2.4$.  At higher redshifts, however,
the galaxies increasingly dominate the ionizing emissivity, producing
essentially all of the ionizing photons just below the Lyman limit at
$z \ge 4$.  These results are consistent with a picture in which
galaxies provide most of the ionizing photons during hydrogen
reionization ($z > 6$), but also indicate that the contribution from
galaxies remains dominant down to much lower redshifts.

The contribution of AGN to the UV background is a subject of ongoing
debate \citep[for a recent discussion see][]{fontanot2012}.
\citet{hm2012}, for example, adopt an AGN ionizing emissivity based on
bolometric luminosity functions compiled by \citet{hopkins2007} that
is roughly a factor of two higher than the \citet{cowie2009}
estimates.  We show the \citet{hm2012} model as a dotted line in
Figure~\ref{fig:eps_with_AGN}.  For these values, AGN are sufficient
to produce essentially all of the ionizing emissivity at $z = 2.4$ and
3.2, though they would still be strongly subdominant at $z \ge 4$.
For this paper we adopt the \citet{cowie2009} estimate of the AGN
emissivity since it is based on direct measurements of the ionizing
flux.  We will argue below that the flat or increasing galaxy
emissivity towards higher redshifts indicates that the efficiency with
which galaxies emit ionizing photons, relative to their non-ionizing
UV output, must increase strongly from $z \sim 3$ to 5.  If we instead
adopt the \citet{hm2012} AGN emissivity, then the evolution in the
galaxy emissivity would be even more pronounced, strengthening this
conclusion.

\begin{figure}
   \begin{center}
   \includegraphics[width=0.45\textwidth]{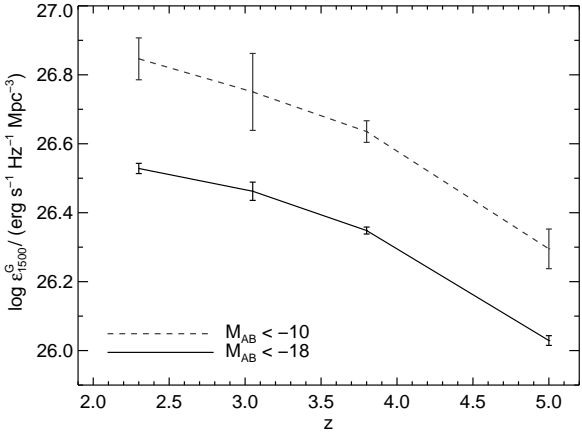}
   \vspace{-0.05in}
   \caption{The non-ionizing specific UV emissivity at 1500~\AA\ from
     dropout-selected galaxies determined from the luminosity
     functions of \citet{reddy2009} ($z = 2.3$ and 3.05) and
     \citet{bouwens2007} ($z = 3.8$ and 5.0). The solid and dashed
     lines show the results when integrating down to $M_{\rm AB} =
     -18$ and $-10$, respectively.  }
   \label{fig:galaxy_eps1500}
   \end{center}
\end{figure}

With an estimate for the ionizing emissivity of galaxies over $2 < z <
5$ in hand, we can now attempt to gain some insight into the evolution
of star-forming galaxies at high redshifts by comparing \epsgal\ to
the integrated galaxy emissivity in the non-ionizing UV continuum,
which essentially traces the unobscured star formation rate density.
For this we use the $\sim$1500~\AA\ rest-frame luminosity functions of
drop-out selected galaxies from \citet{reddy2009} at $z=2.3$ and
$z=3.05$, and from \citet{bouwens2007} at $z=3.80$ and $z=5.0$, each
of which extends significantly over the faint end.  We calculate the
non-ionizing UV emissivity\footnote{In the literature on high-redshift
  galaxies the specific emissivity is sometimes referred to as the UV
  luminosity density, $\rho_{\rm UV}$.}, \epsfifteen, integrated over
$M_{\rm AB} \le -18$ (roughly the observational limit) and $M_{\rm AB}
\le -10$ using the published luminosity function parameters.  Error
estimates for \epsfifteen\ are determined using a Monte Carlo approach
whereby we perturb the binned luminosity functions by their errors,
fit a new luminosity function to the results, and integrate over the
new fit.  The results, which are similar to those obtained by
\citet{robertson2013} at $z=5$, are plotted in
Figure~\ref{fig:galaxy_eps1500}.  The increase when integrating down
to $M_{\rm AB} = -10$ is consistently a factor of $\sim$2 over this
redshift range, due to the fact the faint-end slope remains roughly
constant near $-1.7$.

\begin{figure*}
   \centering
   \begin{minipage}{\textwidth}
   \begin{center}
   \includegraphics[width=0.70\textwidth]{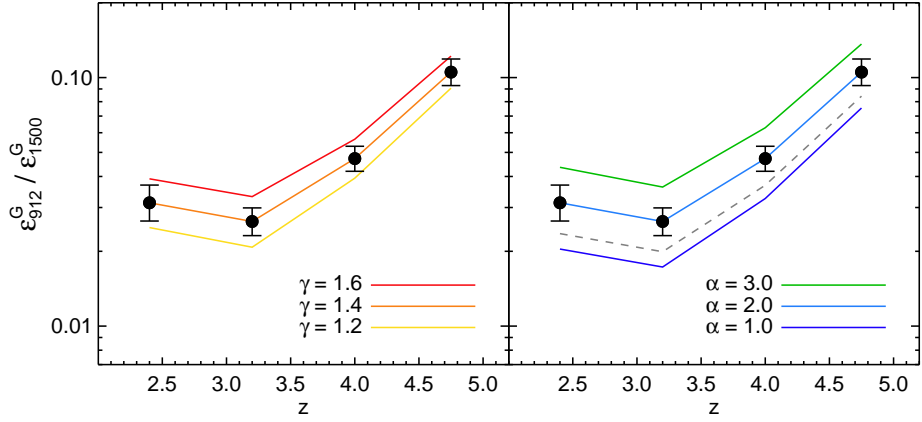}
   \vspace{-0.05in}
   \caption{The ratio of ionizing to non-ionizing specific UV
     emissivity from galaxies.  Filled circles with 1$\sigma$
     statistical error bars show the results for our fiducial
     parameters ($\gamma = 1.4$, $\alpha = 2.0$) and when integrating
     the luminosity density at 1500~\AA\ down to $M_{\rm AB} = -18$.
     In the left-hand panel, solid lines reflect the change in
     $\epsilon^{\rm G}_{912}$ for $\gamma = [1.6,1.4,1.2]$ (top to
     bottom).  In the right-hand panel, solid lines give values for
     $\alpha = [1.0,2.0,3.0]$ (bottom to top).  The dashed line gives
     approximate values for a {\sc starburst99} model with continuous
     star formation and $Z = (1/5)Z_{\odot}$ (see text).}
   \label{fig:eps_ratio}
   \end{center}
   \end{minipage}
\end{figure*}

\begin{figure*}
   \centering
   \begin{minipage}{\textwidth}
   \begin{center}
   \includegraphics[width=0.8\textwidth]{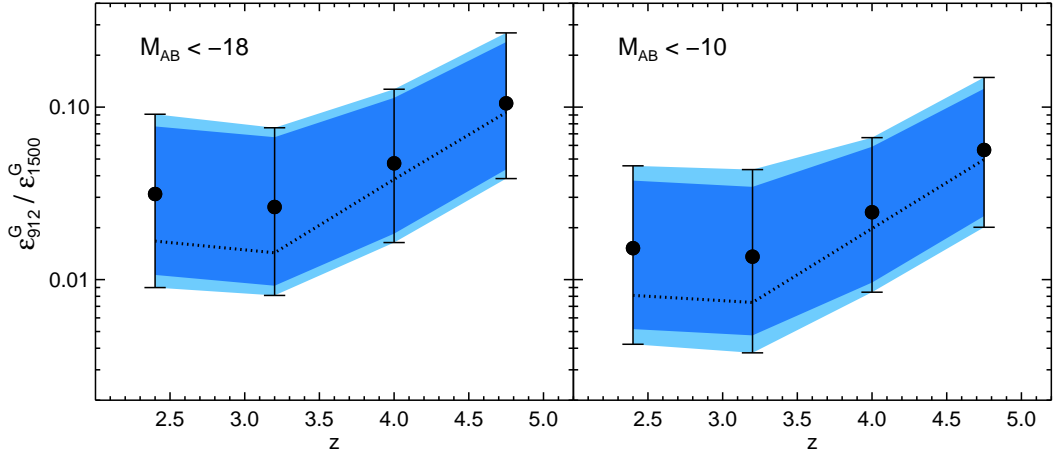}
   \vspace{-0.05in}
   \caption{The ratio of ionizing to non-ionizing specific UV
     emissivity from galaxies, with \epsfifteen integrated down to
     $M_{\rm AB} = -18$ (left-hand panel) and $-10$ (right-hand
     panel).  Filled circles give our results for our fiducial
     parameters ($\gamma=1.4$, $\alpha=2.0$).  The inner shaded band
     gives the total range of systematic uncertainty, while the outer
     shaded band gives the total statistical error.  The dotted lines
     show the fiducial results when \epsgal\ is calculated by
     subtracting the AGN emissivity from \citet{hm2012}.}
   \label{fig:eps_ratio_totalerr}
   \end{center}
   \end{minipage}
\end{figure*}

We compute the ratio of ionizing to non-ionizing galaxy emissivity,
$\epsilon_{912}^{\rm G}/\epsilon_{1500}^{\rm G}$, by interpolating the
\epsfifteen\ values from Figure~\ref{fig:galaxy_eps1500} onto the
redshifts where we measure \epsgal, and again using a Monte Carlo
approach to propagate the errors.  The results are plotted in
Figure~\ref{fig:eps_ratio} for the case where \epsfifteen\ is
integrated down to $M_{\rm AB} = -18$.  For our fiducial parameters
($\gamma=1.4$, $\alpha=2.0$), $\epsilon_{912}^{\rm
  G}/\epsilon_{1500}^{\rm G}$ is flat from $z=2.4$ to 3.2, but then
increases by a factor of $\sim$4 from $z=3.2$ to 4.75.  This change is
driven roughly equally by the decline in \epsfifteen\ and the increase
in \epsgal. Systematic uncertainties are once again considerable,
particularly those related to $\alpha$.  The combined errors are
plotted in Figure~\ref{fig:eps_ratio_totalerr}, where we give the
emissivity ratio when integrating \epsfifteen\ down to both $M_{\rm
  AB} = -18$ and $-10$.  The systematic uncertainties formally permit
$\epsilon_{912}^{\rm G}/\epsilon_{1500}^{\rm G}$ to remain nearly
constant over $2.4 < z < 4.75$.  This requires that $\gamma$ generally
increases with decreasing redshift, a plausible scenario if the
temperature-density relation is somewhat flat at $z \sim 5$.  More
significantly, however, a constant $\epsilon_{912}^{\rm
  G}/\epsilon_{1500}^{\rm G}$ requires that the spectra of the
ionizing sources just below the \hi\ ionization edge become much
softer towards lower redshifts ($\alpha \simeq 1$ at $z=4.75$ to
$\alpha \simeq 3$ at $z=3.2$).  The mean galaxy spectrum may soften
somewhat as stellar populations become more metal rich, but even at
solar metallicity, population synthesis models predict $\alpha
\lesssim 2$ just below 1 Ryd, versus $\alpha \simeq 1.0-1.3$ for $Z
\le (1/5)Z_{\odot}$ \citep{leitherer1999,eldridge2012}.  It therefore
appears likely that the ratio of ionizing to non-ionizing output from
galaxies increases towards higher redshifts.  As noted above, this
trend would be exaggerated (by roughly a factor of two) if we adopted
the AGN emissivity from \citet{hm2012}.  Doing so would decrease
\epsgal\ from galaxies over $2.4 \le z \le 3.2$, while leaving it
largely unchanged at $z = 4.75$.  We show the nominal results for
$\epsilon_{912}^{\rm G}/\epsilon_{1500}^{\rm G}$ under this scenario
as a dotted line in Figure~\ref{fig:eps_ratio_totalerr}.

\section{Discussion}\label{sec:discussion}

\subsection{Comparison with direct measurements of galaxy ionizing emissivity}

We now address whether our estimates of the ionizing emissivity are
compatible with what has been directly observed at high
redshifts. \citet{nestor2013} and \citet{mostardi2013} recently used
narrow-band imaging to measure the ionizing UV luminosities of large
samples of spectroscopically-confirmed $z \sim 3$ color-selected Lyman
break galaxies (LBGs) and narrow band-selected \lya\ emitters (LAEs).
Both studies find the ratio of ionizing to non-ionizing UV luminosity
to be higher for LAEs than for LBGs.  The ratio for an individual
galaxy will depend on a variety of factors, including the underlying
stellar population and the degree to which star-forming regions are
enshrouded in neutral gas, and may not correlate perfectly with
\lya\ emission.  Nevertheless, we can use the results of these studies
to get a rough estimate of the integrated ionizing emissivity from
star-forming galaxies at $z \sim 3$.  Similar calculations were
performed by \citet{nestor2013} and \citet{mostardi2013}, although the
ones presented here will use slightly different assumptions.  For
simplicity, we will primarily adopt the measurements of
\citet{nestor2013}, who find $L_{912}/L_{1500} =
0.056^{+0.039}_{-0.037}$ for LBGs and $L_{912}/L_{1500} =
0.27^{+0.011}_{-0.011}$ for LAEs (inverting their $\eta$ values).  The
values from \citet{mostardi2013} are lower, although consistent within
the error bars.

\citet{nestor2013} find that $\sim$23 per cent of the LBGs in their
sample are LAEs (\lya\ rest equivalent width $\gtrsim$20~\AA),
consistent with previous estimates at $z \sim 3$
\citep{steidel2000,shapley2003,kornei2010}.  This value is also
consistent with trends over $4 < z < 6$ identified by
\citet{stark2011}.  We can obtain a lower limit on the emissivity from
star-forming galaxies by integrating over the \citet{reddy2009}
rest-frame $\sim$1500~\AA\ LBG luminosity function down to $M_{\rm AB}
= -18$ (roughly the observational limit), assuming that 23 per cent of
LBGs are LAEs.  This gives $\epsilon^{\rm SFG}_{912} =
(19.4^{+7.4}_{-7.2}) \times 10^{23}\,{\rm
  erg\,s^{-1}\,Hz^{-1}\,Mpc^{-3}}$, where we have included the
uncertainty in the UV luminosity function as described in
Section~\ref{sec:sources}.  We have so far neglected a possible
luminosity dependence of \lya\ emission \citep{stark2010,stark2011},
but we note that if we assume that zero per cent of LBGs brighter than
$M = -20.25$ (the bright sample of \citet{stark2011}) are LAEs, then
the above value for $\epsilon^{\rm SFG}_{912}$ would only decrease by
16 per cent.  If we integrate down to $M_{\rm AB} = -10$ and assume
that 23 per cent of LBGs are LAEs at all luminosities, then we find
$\epsilon^{\rm SFG}_{912} = (38.1^{+18.4}_{-15.8}) \times
10^{23}\,{\rm erg\,s^{-1}\,Hz^{-1}\,Mpc^{-3}}$.  A reasonable upper
limit can be obtained by instead assuming that all LBGs fainter than
$M_{\rm AB} = -18$ are LAEs, in which case $\epsilon^{\rm SFG}_{912} =
(67.7^{+37.1}_{-31.7}) \times 10^{23}\,{\rm
  erg\,s^{-1}\,Hz^{-1}\,Mpc^{-3}}$.  This value becomes $\epsilon^{\rm
  SFG}_{912} = (33.8^{+34.7}_{-15.9}) \times 10^{23}\,{\rm
  erg\,s^{-1}\,Hz^{-1}\,Mpc^{-3}}$ if we adopt the ionizing to
non-ionizing UV luminosity ratios from \citet{mostardi2013}.  We plot
the extremes of these emissivity values within the 68 per cent error
ranges in Figure~\ref{fig:eps_with_AGN}, noting that the lower limit
on $\epsilon^{\rm SFG}_{912}$ is highly conservative as it neglects
any contribution from galaxies fainter than $\sim$0.1$L^{*}$.  The
overlap with our indirect estimate of the total emissivity from
galaxies is very good, indicating that star-forming galaxies can
indeed provide the majority of ionizing photons in the IGM at $z \sim
3$.

\subsection{Redshift evolution in the ionizing efficiency of galaxies}

Our results for $\epsilon_{912}^{\rm G}/\epsilon_{1500}^{\rm G}$
suggest that the luminosity-weighted ionizing ``efficiency'' of
galaxies may be increasing with redshift from $z \sim 2-3$ to $z \sim
5$ (Figure~\ref{fig:eps_ratio_totalerr}).  We reiterate that while our
measurement of \epsgal\ includes ionizing photons from all galaxies,
the non-ionizing galaxy emissivity is only directly measured from
galaxies brighter than $M_{\rm AB} \simeq -18$.  The apparent increase
in $\epsilon_{912}^{\rm G}/\epsilon_{1500}^{\rm G}$ from $z=3.2$ to
$z=4.75$ could therefore be partially negated if the relative
contribution to \epsfifteen\ from fainter galaxies increases in a way
not captured by the measured UV luminosity functions.  The faint end
slope of the UV luminosity function down to $M_{\rm AB} = -18$ has
been found to remain nearly constant over this redshift interval, so
such a trend would require that the luminosity functions deviate
substantially from Schechter functions at fainter luminosities.  No
such departure has be been seen down to $M_{\rm AB} = -13$ in
observations of lensed galaxies at $z \sim 2$ \citep{alavi2013}.  The
shape of the luminosity function for faint sources could be different
at higher redshifts; however, at face value there appears to be a real
evolution in the efficiency with which galaxies produce and/or emit
ionizing photons relative to their total non-ionizing UV output.

Other trends in the properties of LBGs consistent with such an
evolution have been noted.  \citet{shapley2003} and \citet{jones2012}
find a trend of decreasing absorption from low-ionization metals with
increasing \lya\ equivalent width in LBGs at $z \sim3-4$.  The
decrease in absorption from low-ionization metals appears to be driven
by a decrease in \hi\ covering fraction \citep{jones2013}, consistent
with the correlation of \lya\ emission and ionizing flux found by
\citet{nestor2013}.  Together with the observation from
\citet{stark2010,stark2011} that the fraction of \lya\ emitters among
LBGs becomes larger with redshift over $4 < z < 6$, these trends
suggests that the integrated ratio of ionizing to non-ionizing flux
from LBGs should indeed be increasing with redshift, as we find.

We note that our values of $\epsilon_{912}^{\rm
  G}/\epsilon_{1500}^{\rm G}$ do not necessarily require implausibly
high escape fractions of ionizing photons.  The {\sc starburst99}
model mentioned above, which had $t_{\rm age} = 500$~Myr, constant
star formation rate, and $Z = (1/5)Z_{\odot}$, has an intrinsic
luminosity ratio of $L_{\nu}^{912}/L_{\nu}^{1500} \simeq 0.15$.  For
our fiducial value of $\gamma$, such a spectrum would have
$\epsilon_{912}^{\rm G}/\epsilon_{1500}^{\rm G} = 0.045$ at $z=4.75$
when integrating \epsfifteen\ down to $M_{\rm AB} = -10$.  The
luminosity-weighted mean relative escape fraction would thus be
$f_{\rm esc}^{912}/f_{\rm esc}^{1500} \simeq 0.30$.  However, a {\sc
  bpass} model stellar population \citep{eldridge2012} with the same
parameters but including massive binaries and quasi-homogeneous
evolution of rapidly rotating stars has a similar shape in the
ionizing region of the spectrum as the {\sc starburst99} model but a
larger intrinsic luminosity ratio, $L_{\nu}^{912}/L_{\nu}^{1500}
\simeq 0.29$.  In this case the mean relative escape fraction would be
$f_{\rm esc}^{912}/f_{\rm esc}^{1500} \simeq 0.16$, and somewhat lower
if $\gamma$ is less than our fiducial value.

\subsection{Implications for Hydrogen Reionization}

The analysis above provides at least two insights relevant to hydrogen
reionization.  First, the redshift evolution in $\epsilon_{912}^{\rm
  G}/\epsilon_{1500}^{\rm G}$ indicates that, for a given non-ionizing
emissivity, galaxies at $z \sim 5$ emit more ionizing photons than
their $z \sim 2-3$ counterparts.  Even if this trend flattens at $z >
5$, this suggests that the galaxies responsible for hydrogen
reionization produce and/or emit ionizing photons more
``efficiently'', in a luminosity-weighted sense, than more
well-studied galaxies at lower redshifts.  Reionization models based
on observed UV luminosity functions at $z > 6$ have generally
concluded that such an enhanced efficiency is required for galaxies to
complete reionization by $z = 6$, and have often invoked a high escape
fraction for reionization-era galaxies
\citep[e.g.,][]{ouchi2009,kuhlen2012,mitra2013,robertson2013}.  Here
we present empirical evidence that such a high efficiency, whether due
to a changing escape fraction or a combination of factors, is well
motivated by long-term trends in $\epsilon_{912}^{\rm
  G}/\epsilon_{1500}^{\rm G}$ observed in the post-reionization epoch.
We note that our fiducial increase in $\epsilon_{912}^{\rm
  G}/\epsilon_{1500}^{\rm G}$ of a factor of $\sim$4 from $z=3.2$ to
4.75 is of similar magnitude to the increase in the escape fraction
adopted by \cite{hm2012}, who assume a factor of $\sim$3 increase over
the same redshift interval.

Second, the ionizing emissivity at $z \sim 4-5$ is around a factor of
two larger than previous measurements have indicated.  At $z = 4.75$,
we find $\dot{N}_{\rm ion} = 2$ to 14 ionizing photons per atom per
gigayear.  This higher emissivity, if it extends much beyond $z \sim
5$ (but see \citet{bolton2007}), would make it easier to explain how the
last and potentially densest 10 percent of the IGM is ionized in the
170 million years between $z=7$ and $z=6$
\citep{mortlock2011,bolton2011b}, alleviating the tension between this
measurement and constraints on the reionization history from
independent observational probes \citep[e.g.,][]{ciardi2012,jensen2013}.

\subsection{The Evolution of the UV background at \boldmath{$z > 5$}}

\begin{figure}
   \begin{center}
   \includegraphics[width=0.45\textwidth]{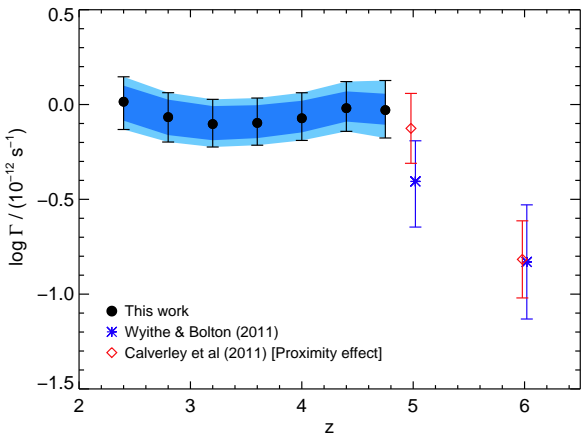}
   \vspace{-0.05in}
   \caption{The evolution in the \hi\ ionization rate over $2 < z <
     6$.  Circles with error bars are the present results, with
     systematic and statistical uncertainties given by the inner and
     outer shaded regions, respectively.  Literature values at $z \ge
     5$ are from \citet{wyithe2011} (asterisks) and
     \citet{calverley2011} (diamonds).}
   \label{fig:G12_with_zgt5}
   \end{center}
\end{figure}

Finally, we turn towards comparing our results for the UV background
to measurements at $z > 5$.  Similar to previous works, we find that
the \hi\ ionization rate remains fairly flat over $2.4 \le z \le
4.75$.  In contrast, \G\ has been found to be substantially lower at
$z \sim 6$ based on measurements using both the mean \lya\ opacity
\citep{bolton2007,wyithe2011} and the quasar proximity effect
\citep{calverley2011}.  We plot the redshift evolution of \G\ over $2
< z < 6$ in Figure~\ref{fig:G12_with_zgt5}, where we have slightly
adjusted the \citet{wyithe2011} results to be consistent with our
fiducial value of $\gamma=1.4$ using the scaling from
\citet{bolton2007}.  Comparing our results at $z = 4.75$ to the
weighted average of the literature results at $z=6$, \G\ appears to
increase by $0.8 \pm 0.3$ dex over $\Delta z \sim 1$.

As discussed above, a change in \G\ may be due to a change either in
the ionizing emissivity or in the opacity of the IGM to ionizing
photons (i.e., the mean free path).  The fact that \eps\ is remains
roughly constant over $2 < z < 5$, and may even increase with redshift
over $3 < z < 5$, suggests that a factor of $\gtrsim$4 decrease in the
ionizing emissivity from $z \sim 5$ to $z \sim 6$ is unlikely.  The
non-ionizing galaxy UV emissivity, integrated down to $M_{\rm AB}=
-10$, decreases by at most a factor of 1.6 \citep{robertson2013}.
This admittedly assumes that the luminosity function remains
Schechter-like over the entire faint end.  At face value, however,
unless the ratio of ionizing to non-ionizing emissivity decreases with
redshift over $5 < z < 6$, a trend contrary that found over $3 < z <
5$, this supports the conclusion that the decline in \G\ is unlikely
to be driven solely by a decrease in the ionizing emissivity.  The
rapid evolution in the UV background over $5 < z < 6$ may therefore
indicate a substantial change in the mean free path.  Extrapolating
our fit to the effective \hi\ column density distribution over $2 < z
< 5$, we would expect a factor of $\sim$2 decline in \mfp\ between
$z=5$ and 6.  A break in the mean free path evolution would therefore
be required to explain the decline in \G\ if the emissivity remained
constant or increased with redshifts.  Such a deviation in the
evolution of \mfp\ may already be evident in the $z \simeq 5.1$
measurement of Worseck et al. (in prep), which was not included in our
fit to $f(N_{\rm H\,I},z)$ because it is outside the redshift range of
our analysis (see Figure~\ref{fig:cddf}).  A factor of two decrease in
\mfp\ coupled with a decline in \eps\ proportional to that in
\epsfifteen\ would also marginally be able to explain the drop in \G.

A rapid evolution in the mean free path over $5 < z < 6$, if present,
would be consistent with expectations for the `post-overlap' phase of
reionization.  As pointed out by \citet{me2000} and \citet{furoh2005},
once the IGM is essentially fully ionized, the evolution in \G\ is
driven by the photo-evaporation of optically thick Lyman-limit
systems, which largely regulate the mean free path.  If such a process
is occurring over $5 < z < 6$, then this would be consistent with a
picture in which the overlap phase of reionization ends not long
before $z \sim 6$, as suggested by the $\gtrsim$10 percent neutral
fraction measured around the $z = 7.1$ quasar ULAS~J1120$+$0641
\citep{mortlock2011,bolton2011b}.  On the other hand,
\citet{mcquinn2011b} have noted that a small change in the ionizing
emissivity can potentially produce a large change in \G, particularly
at $z \sim 5-6$, due to the coupling between \G, \eps, and \mfp.  The
increase in \G\ from $z \sim 6$ to 5 could therefore simply reflect a
modest increase in \eps\ over this interval, consistent with the
observed evolution in \epsfifteen, although other factors would be
required to explain why \G\ then remains essentially flat over $ 2 < z
< 5$.  

Finally, we caution here that \G\ at $z = 6$ may be underestimated if
there is significant scatter in the intensity of the UVB at this
redshift.  \citet{bolton2007} note that the photoionization rate may be
underestimated by up to 10 per cent at $z\sim 6$ due to the influence
of UVB fluctuations, although they incorporate this uncertainty into
their published \G\ measurements.  \citet{mesinger2009}
also note UVB fluctuations will alter the inferred photoionization
rates by a few per cent at these redshifts.

\section{Summary}\label{sec:summary}

We have presented new measurements of the intensity of the UV
background and the global ionizing emissivity over $2 < z < 5$.  The
results are based on recent measurements of the mean \lya\ opacity and
temperature of the IGM, as well as the opacity of the IGM to ionizing
photons.  This study benefits from the precision of these new
measurements, as well as from the fact that the measurements are
self-consistent over a wide redshift range.

Similar to previous works
\citep[e.g.,][]{bolton2005,bolton2007,fg2008c}, we find that the
hydrogen photoionization rate, \G, stays roughly constant over $2 < z
< 5$.  We find a factor of two larger \G\ than that obtained by
\citet{fg2008c}, however, a difference due mainly to our lower IGM
temperatures.  Our results for the ionizing emissivity are also
considerably higher than previous estimates
\citep{bolton2007,kuhlen2012}.  This is due to a combination of lower
IGM temperatures, a higher opacity to ionizing photons, and the fact
that our calculations take into account cosmological radiative
transfer effects.

We calculate the emissivity of star-forming galaxies by subtracting
estimates for the emissivity of AGN made by \citet{cowie2009} from our
total emissivity measurements.  We find that galaxies dominate the UV
background near 1 Ryd, especially at $z \ge 4$, and that \epsgal\ may
increase with redshift over this interval.  Our estimate for
\epsgal\ is consistent with estimates of the emissivity from
star-forming galaxies at $z \sim 3$ based on direct measurements of
the ionizing output from Lyman break galaxies and \lya\ emitters
\citep{nestor2013,mostardi2013}.  Comparing \epsgal\ to the
non-ionizing UV emissivity of galaxies over $2 < z < 5$ determined
from rest-frame UV luminosity functions \citep{reddy2009,bouwens2007},
we find that the ionizing ``efficiency'' of galaxies remains
relatively flat over $2.4 \le z \le 3.2$ but may increase
substantially from $z = 3.2$ to $z = 4.75$.  Assuming that the faint
end of the luminosity function over this interval is well described by
the measured Schechter functions, the ratio of ionizing to
non-ionizing emissivity would remain flat only if the ionizing source
spectrum near 912~\AA\ softened dramatically from $z \sim 5$ to $z
\sim 3$.  We note that the increase in $\epsilon^{\rm
  G}_{912}/\epsilon^{\rm G}_{1500}$ with redshift would become even
stronger if we used the higher AGN emissivity adopted by
\citet{hm2012}, which is based on the AGN luminosity functions of
\citet{hopkins2007}.

Our results carry multiple implications for hydrogen reionization.
First, we have shown that the ionizing efficiency of galaxies appears
to increase with redshift in the post-reionization epoch.  This
evolution supports the conclusion often drawn from reionization
modeling that reionization-era galaxies must produce and/or emit
ionizing photons significantly more efficiently than their
lower-redshift counterparts in order to complete reionization by $z =
6$ \citep[e.g.,][]{ouchi2009,kuhlen2012,mitra2013,robertson2013}.
Second, we find that ionizing photons appear to be a factor of two
more abundant in the IGM at $z \sim 5$ compared to previous studies,
with $\dot{N}_{\rm ion} \simeq 2-14$ ionizing photons/atom/gigayear
($\dot{N}_{\rm ion} \sim 5$ photons/atom/gigayear for realistic galaxy
spectra) .  This high emissivity, if it persists to higher redshifts,
would explain how the last, and potentially densest, 10 per cent of
the IGM may have been reionized after $z = 7$
\citep{mortlock2011,bolton2011b}.

Finally, the relatively flat ionization rate we find over $2.4 \le z
\le 4.75$ contrasts with the sharp decline measured from $z \sim 5$ to
6.  The low measurement of \G\ at $z \sim 6$ has previously been
interpreted as evidence for a low ionizing emissivity and a
``photon-starved'' mode of reionization
\citep{bolton2007,calverley2011}.  In light of the fact that the
emissivity we measure appears to remain flat or increase with redshift
over $3 < z < 5$, the decrease in \G\ may instead signify a rapid
evolution in the mean free path, consistent with the photoionization
of Lyman limit systems at the end of the `post-overlap' phase of
reionization \citep[e.g.,][but see McQuinn et al. 2011]{furoh2005}.
Under this scenario, ionizing photons may have been considerably more
abundant during the final stages of reionization than has previously
been recognized.

\section*{Acknowledgements} We are extremely grateful to Michele Fumagalli 
and Gabor Worseck, who allowed us to use their mean free path measurements 
ahead of publication.  We are also grateful to John Eldridge, Martin Haehnelt,
John O'Meara, Xavier Prochaska, and Naveen Reddy for helpful
conversations.  We thank Volker Springel for making GADGET-3
available.  We also thank the anonymous referee for helpful comments.  
The hydrodynamical simulations used in this work were
performed using the Darwin Supercomputer of the University of
Cambridge High Performance Computing Service
(http://www.hpc.cam.ac.uk/), provided by Dell Inc. using Strategic
Research Infrastructure Funding from the Higher Education Fund- ing
Council for England.  GDB gratefully acknowledges support from the
Kavli Foundation.  JSB acknowledges the support of a Royal Society
University Research Fellowship.

\bibliographystyle{apj} \bibliography{uvb_refs}

\appendix

\section{Scaling relations for ${\mathbf \Gamma}$}\label{app:scaling}

\begin{figure}
   \begin{center}
   \includegraphics[width=0.45\textwidth]{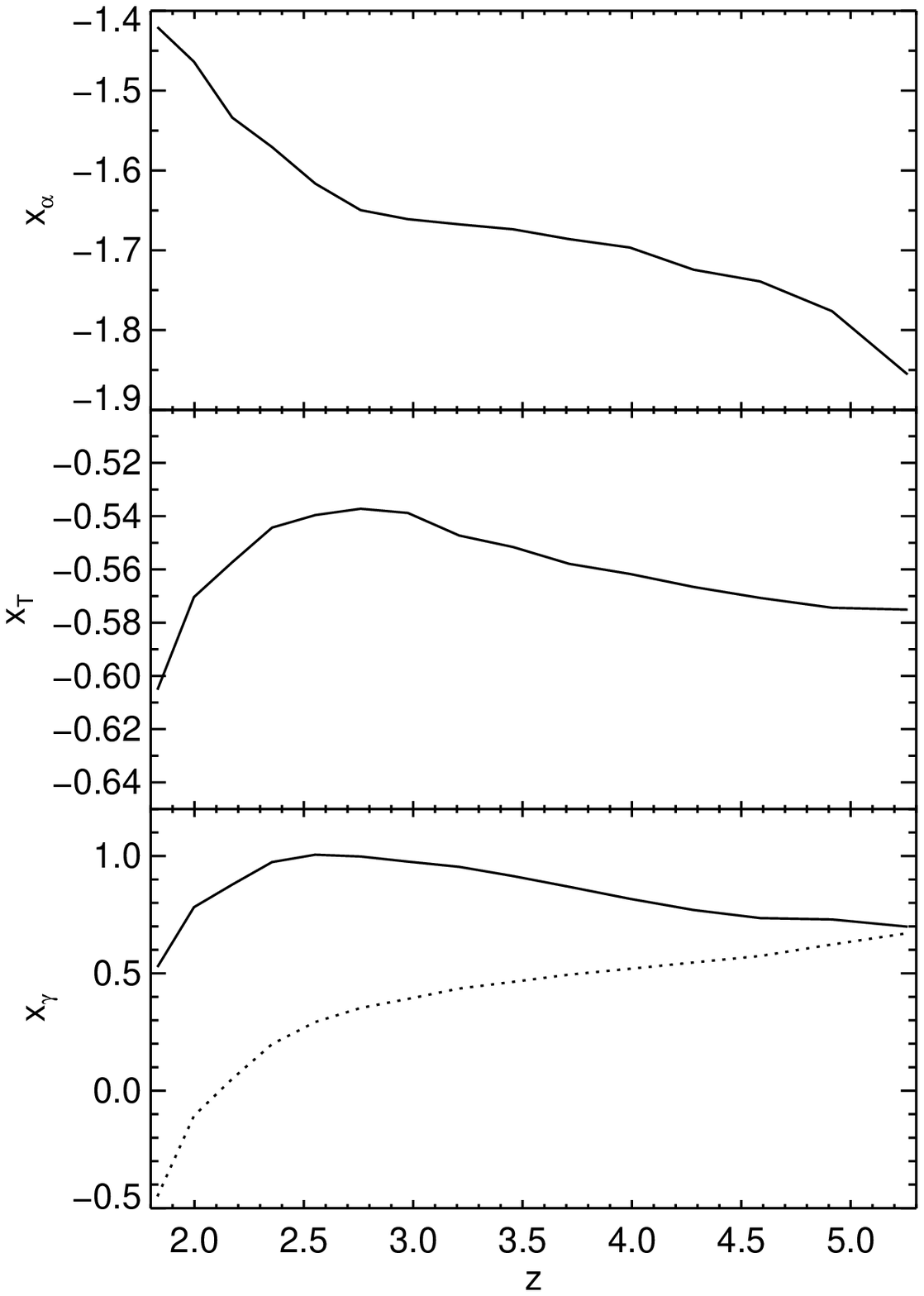}
   \vspace{-0.05in}
   \caption{Scaling coefficients for \G\ as a function of redshift.
     $x_{\alpha}$ (upper panel) and $x_{\rm T}$ (middle panel) are the
     power law indices for the scaling of \G\ with \taueffa\ and
     $T(\bar{\Delta})$, respectively.  $x_{\gamma}$ (lower panel) is
     the exponential coefficient for the scaling of \G\ with $\gamma$
     in the cases where $T(\bar{\Delta})$ and $T_0$ are held fixed
     (solid and dotted lines, respectively).}
   \label{fig:coeffs}
   \end{center}
\end{figure}

Here we update the scaling relations from \citet{bolton2005} and
\citet{bolton2007} describing the dependence of the photoionization
rate on the effective optical depths and the parameters of the
temperature-density relation.  For the effective optical depth, power
law relations of the form $\Gamma \propto (\tau_{\rm
  eff}^{\alpha})^{x_{\alpha}}$ were empirically determined as a
function of redshift from the simulations.  These were found to
describe the dependence of \G\ on \taueffa\ very accurately, with
$x_{\alpha}$ decreasing from -1.46 at $z = 1.998$ to -1.78 at $z =
4.915$ (Figure~\ref{fig:coeffs}), similar to the results of
\citet{bolton2005} and \citet{bolton2007}

In order to assess the scaling of \G\ with gas temperature, we imposed
power law temperature-density relations of the form $T(\Delta) =
\bar{T}(\Delta/\bar{\Delta})^{\gamma-1}$ on our fiducial simulation,
and separately evaluated the dependence of \G\ on \Tbar\ and $\gamma$.
When changing the gas temperature, the hydrogen neutral fraction was
also scaled as $f_{\rm H\,I} \propto T^{-0.72}$.  Changes in \G\ with
\Tbar\ were found to be well described by power laws of the form
$\Gamma \propto \bar{T}^{x_{\rm T}}$, with ${x_{\rm T}} \simeq 0.55$
but varying slightly with redshift (Figure~\ref{fig:coeffs}).  This
scaling is similar to the results obtained by \citet{bolton2005} and
\citet{bolton2007}.  The slightly weaker dependence of \G\ on
\Tbar\ compared to what one would expect from
equation~(\ref{eq:tau_lya}) reflects the fact that the decrease in
\taueffa\ with \G\ due to the decrease in the recombination rate is
somewhat offset by an increase in \taueffa\ due to increased thermal
broadening.

We evaluated the dependence of \G\ on the slope of the
temperature-density relation by varying $\gamma$ while holding
\Tbar\ fixed.  The result was a an exponential scaling of the form
$\Gamma \propto \exp{(x_{\gamma}\gamma)}$, with $x_{\gamma}$ varying
with redshift between 0.76 and 1.00 (Figure~\ref{fig:coeffs}).  This
is a stronger dependence on $\gamma$ than if the temperature was
always held fixed at the mean density, the results for which we also
plot in Figure~\ref{fig:coeffs}.  This difference is simply due to the
fact that \taueffa\ is mainly sensitive to the opacity of the voids.
We note that \G\ becomes largely insensitive to $\gamma$ if the
temperature is known near $\Delta \sim 0.5$ (although the precise
value of $\Delta$ depends on the redshift).  In principle, this means
that if \G\ can be accurately measured by some other method, such as
the proximity effect, it may be possible to infer a value for
$\gamma$.

\section{Impact of Jeans Smoothing}\label{app:jeans}

\begin{figure}
   \begin{center}
   \includegraphics[width=0.45\textwidth]{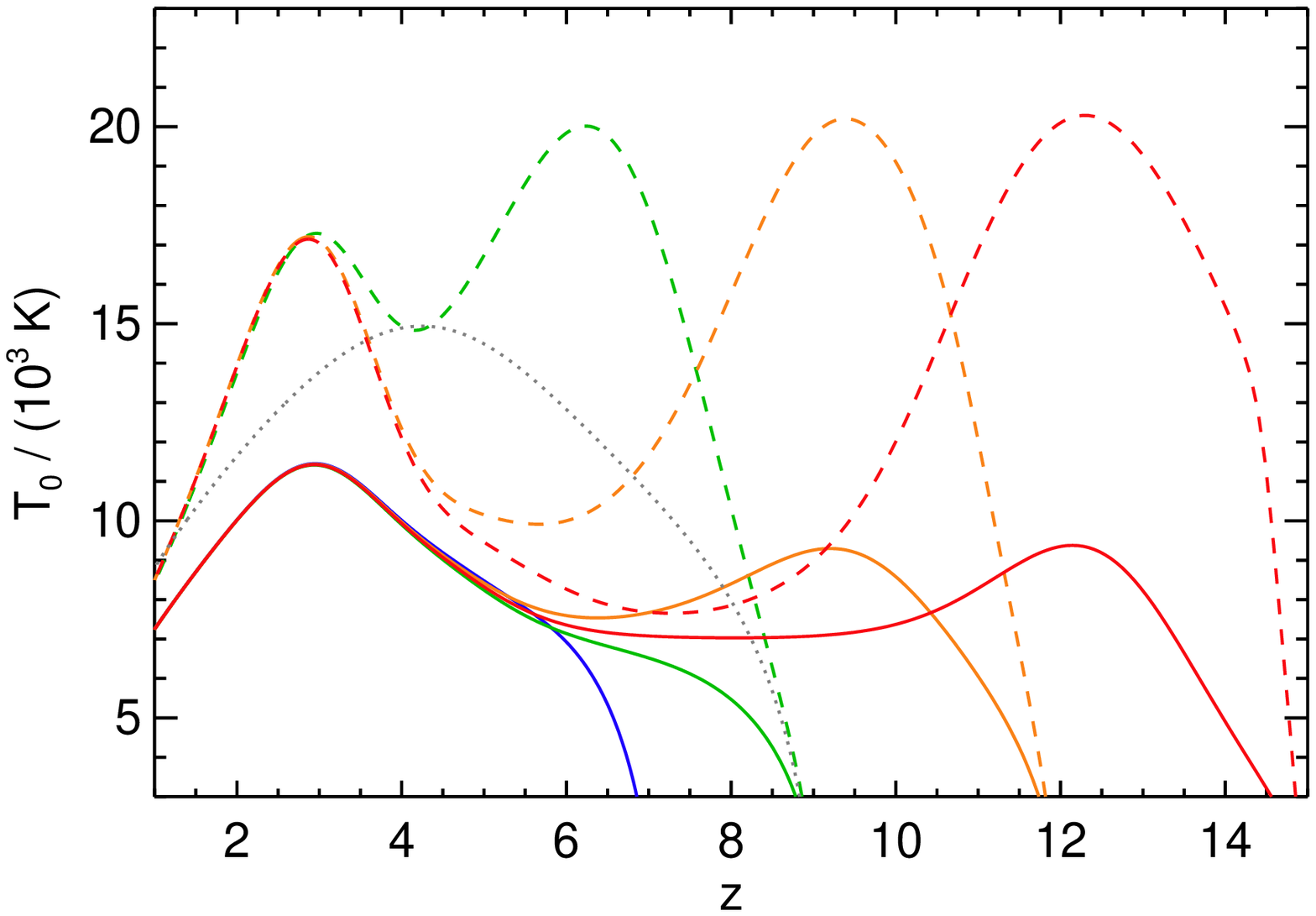}
   \vspace{-0.05in}
   \caption{Thermal histories for the hydrodynamical simulations used
     in this work.  Temperatures at the mean density are plotted as a
     function of redshift.  The solid and dashed lines show runs used
     to assess the impact of Jeans smoothing on \G.  The dotted line
     shows the thermal history for the C15 and R runs.}
   \label{fig:Thist}
   \end{center}
\end{figure}

In addition to the role of gas temperature in setting the
\hi\ recombination rate and the thermal broadening of absorption
lines, the integrated thermal history of the IGM can potentially
affect our estimate of \G\ through its impact on the baryon density
distribution.  On large scales the baryons will generally trace the
underlying dark matter, whose density field can be predicted from
cosmology.  On small scales, however, the baryon distribution will be
smoothed due to thermal pressure.  This effect, known as ``Jeans
smoothing'', is particularly significant during hydrogen reionization,
when the temperature of the gas is increased by orders of magnitude
\citep[e.g.,][]{pawlik2009}.

To test whether Jean smoothing may impact our \G\ measurements, we ran
a set of simulations in which the timing and amplitude of
photoionization heating during hydrogen (as well as at lower
redshifts) were varied.  The thermal histories for these runs are
shown in Figure~\ref{fig:Thist}.  Our fiducial values of \G\ were
calculated using the density and peculiar velocity fields from the run
in which hydrogen reionization begins at $z = 12$ and the gas is
heated to $\sim$9,000 K by $z \simeq 9$.  Using the density and
peculiar velocity fields from other runs was found to have a minimal
($\lesssim$0.02 dex) impact on \G.  Uncertainties related to Jeans
smoothing are nevertheless included in our systematic error budget
(Tables~\ref{tab:Gamma_errors} and \ref{tab:emissivity_errors}).

\section{Emissivity Calculation}\label{app:em_calc}

\begin{figure}
   \begin{center}
   \includegraphics[width=0.45\textwidth]{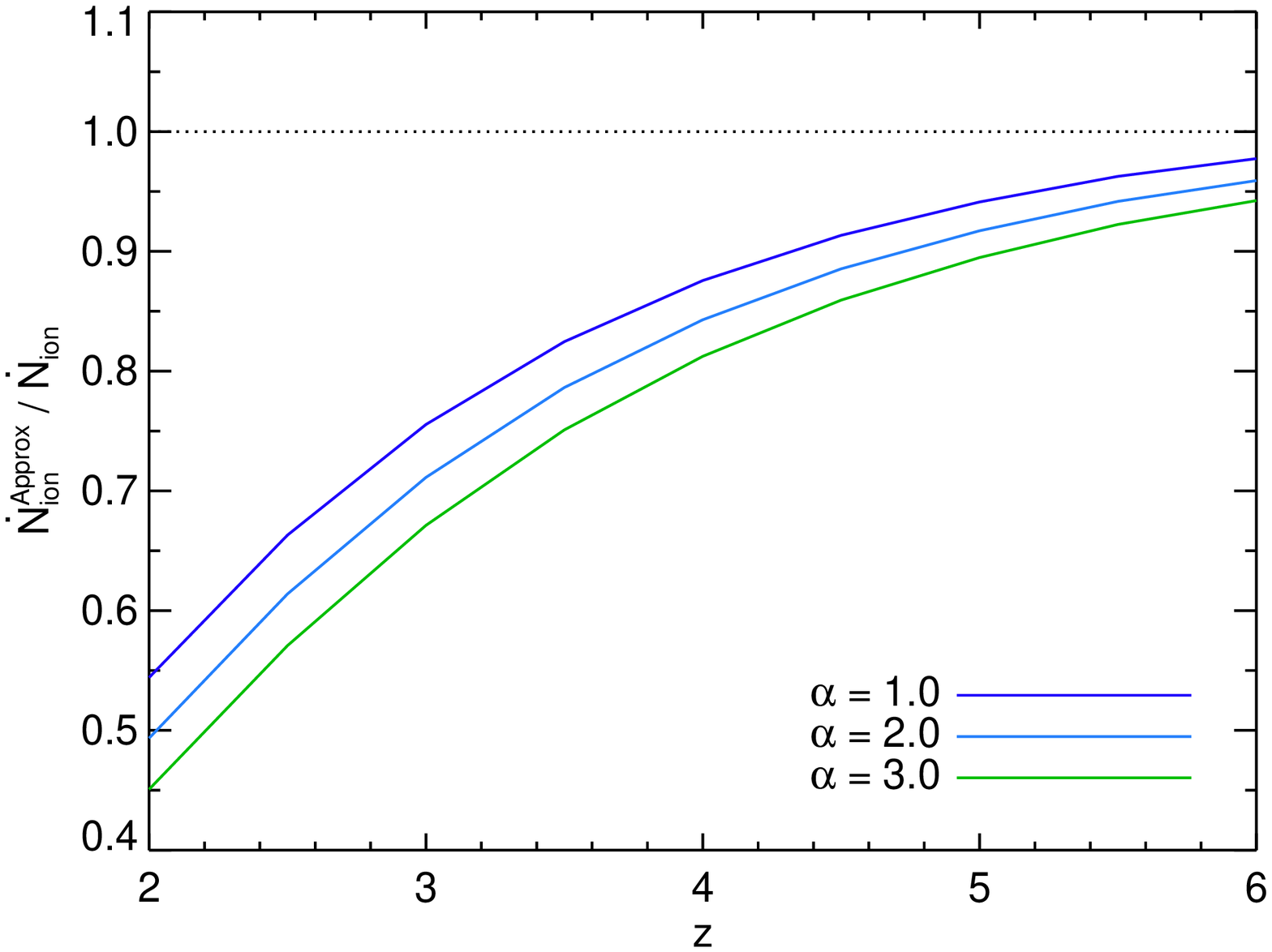}
   \vspace{-0.05in}
   \caption{The ratio of the ionizing emissivity computed from the
     approximation given by equation~(\ref{eq:Nion_approx}) to that
     computed using full cosmological radiative transfer.  Solid lines
     are for $\alpha = 1.0$, 2.0 and 3.0 (top to bottom).}
   \label{fig:emissivity_approx}
   \end{center}
\end{figure}

In Section~\ref{sec:emissivity} we described how we compute the
ionizing emissivity from the \hi\ ionization rate and the opacity of
the IGM+CGM to ionizing photons.  Our calculations were based on the
full equation of cosmological radiative transfer, which allowed us to
take into account the redshifting of ionizing photons as they
propagate through the IGM \citep[e.g.,][]{hm1996}.

It has been common to simplify the calculation of the emissivity by
assuming that the mean free path of ionizing photons is short enough
to neglect the redshifting effect.  Under this ``local source''
approximation, equation~(\ref{eq:specific_intensity}) reduces to
\citep[e.g.,][]{schirber2003}
\begin{equation}\label{eq:specific_intensity_approx}
J_{\nu} \approx \frac{1}{4\pi}\lambda(\nu,z)\epsilon_{\nu}(z) \, .
\end{equation}
In addition, calculations of the mean free path based on the
\hi\ column density distribution (equation~(\ref{eq:cddf})) have also
commonly used an approximate expression that neglects redshifting
(i.e., $z_{0} \simeq z$), such that
\begin{eqnarray}\nonumber\label{eq:mfp_approx}
\lambda(\nu,z) & \approx & \frac{(\beta_{N}-1)c}{\Gamma(2-\beta_{N})N_{\rm LL}\sigma_{912}^{\beta_{N}-1}}
   \left( \frac{\nu}{\nu_{912}} \right)^{2.75(\beta_{N}-1)} \\ & & \times \frac{1}{(1+z)^{\beta_{z}+1}H(z)}
\end{eqnarray}
\cite[e.g.,][equation (22)]{madau1999,fg2008c}, where $\Gamma$ here is
the gamma function.

Combining equations~(\ref{eq:Gamma}), (\ref{eq:Nion}) and
(\ref{eq:specific_intensity_approx}), and integrating over frequency
gives \citep[e.g.,][]{kuhlen2012}
\begin{equation}\label{eq:Nion_approx}
\dot{N}_{\rm ion}(z) \approx \frac{1}{(1+z)^3} \frac{\Gamma(z)}{\sigma_{912}\lambda_{912}(z)} \frac{(\alpha_{\rm bg} + 2.75)}{\alpha} \, .
\end{equation}
Here we have modeled the specific intensity as a power law in
frequency, $J_{\nu} = J_{912} (\nu/\nu_{912})^{\alpha_{\rm bg}}$ for
$\nu \ge \nu_{912}$, where for a power-law distribution of \hi\ column
densities, $f(N_{\rm H\,I}) \propto N_{\rm H\,I}^{-\beta_{N}}$, the
slope of the ionizing background is given by $\alpha_{\rm bg} = \alpha
- 2.75(\beta_{N}-1)$.

In Figure~\ref{fig:emissivity_approx} we plot the approximate value of
\Nion\ given by equations~(\ref{eq:mfp_approx}) and
(\ref{eq:Nion_approx}) as a fraction of the value obtained from the
full calculation described in Section~\ref{sec:emissivity}.  For this
comparison we assume that \G\ is constant with redshift, and adopt our
nominal \hi\ column density distribution parameters $[A,\beta_{\rm
    N},\beta_{z}] = [0.93,1.33,1.92]$.  The approximate expression for
\Nion\ asymptotically approaches the exact value with increasing
redshift, with $\dot{N}_{\rm ion}^{\rm approx}/\dot{N}_{\rm ion} \ge
0.8$ for $z > 3.6$ and $\alpha < 2$.  At $z \sim 2-3$, however, the
approximate expression is significantly too low, up to a factor of two
at $z = 2$.  This can be understood from the fact that, once the mean
free path becomes sufficiently large, a significant fraction of
ionizing photons will redshift beyond the Lyman limit before they are
absorbed.  The approximate expression for \Nion\ neglects this effect,
and so requires a lower emissivity for a given ionization rate.  The
effect is greater for softer ionizing spectra, where a larger fraction
of the ionizing flux is concentrated near the Lyman limit.  Even for
$\alpha = 1.0$, however, the approximate calculation delivers a value
for \Nion\ that is significantly too low at $z < 4$.

Finally, we emphasize that the results plotted in
Figure~\ref{fig:emissivity_approx} apply only when the approximate
expressions for both \mfp\ and \Nion\ are used.  If
equation~\ref{eq:Nion_approx} is instead evaluated using the exact
\mfp\ values calculated from $f(N_{\rm H\,I},z)$ as described in
Section~\ref{sec:emissivity_method}, or measured directly from
composite spectra \citep[][Worseck et al.,
  in prep]{prochaska2009b,omeara2013,fumagalli2013}, the emissivity will be
underestimated by an even greater factor.

\end{document}